\newcommand{\CLASSINPUTbottomtextmargin}{1.0in}
\newcommand{\CLASSINPUTtoptextmargin}{0.8in}
\documentclass[lettersize,journal]{IEEEtran}

\usepackage{fancyhdr}
\usepackage{epsfig}
\usepackage{threeparttable}
\usepackage{epsf,epsfig}
\usepackage{amsthm}
\usepackage[cmex10]{amsmath}
\usepackage{amssymb, amsfonts}
\usepackage[noadjust]{cite}
\usepackage{dsfont}
\usepackage{subfigure}
\usepackage{color}
\usepackage{soul}
\usepackage{amssymb}
\usepackage{accents}
\usepackage{algorithm}
\usepackage{diagbox}
\usepackage[english]{babel}
\usepackage{url}
\usepackage{framed}
\usepackage{multirow}
\usepackage{graphicx}
\usepackage{indentfirst}
\usepackage{url}
\usepackage{framed} 
\usepackage[colorlinks,
linkcolor=black,
anchorcolor=black,
citecolor=black]{hyperref}

\newtheorem{proposition}{Proposition}

\newtheorem{remark}{Remark}
\def\phi{\varphi}

\def\SNR{\mathsf{SNR}}

\def\({\left(}
\def\){\right)}

\def\b0{{\mathbf{0}}}

\allowdisplaybreaks[2]

\def\bf#1{\mathbf{#1}}

\def\rv#1{\textcolor{black}{#1}}

\usepackage{mathrsfs}
\usepackage{algorithm}
\usepackage{algorithmic}
\usepackage[justification=centering]{caption}
\begin{document}	
	\pagestyle{empty}
	\title{\huge Beamforming Design for Semantic-Bit Coexisting Communication System }
	\author{Maojun Zhang, Guangxu Zhu, Richeng Jin, Xiaoming Chen, Qingjiang Shi, Caijun Zhong, Kaibin Huang
	\thanks{
	M. Zhang, R. Jin, Xiaoming Chen, and C. Zhong 
	is with the Department of Information and Communication Engineering, Zhejiang University, Hangzhou, China, 310007, and also with Zhejiang Provincial Key Lab of Information Processing, Communication, and Networking (IPCAN), Hangzhou, China, 310007
	(Email: $\{$zhmj, richengjin, chen\_xiaoming, caijunzhong$\}$@zju.edu.cn.). 
	G. Zhu is with Shenzhen Research Institute of Big Data, Shenzhen, China (Email: gxzhu@sribd.cn). 
	Q. Shi is with the School of Software Engineering, Tongji University, Shanghai 200092, China, and also with Shenzhen Research Institute of Big Data, Shenzhen 518172, China (email: shiqj@tongji.edu.cn).
	K. Huang is with the University of Hong Kong, Hong Kong (email: huangkb@eee.hku.hk).

	}	
	\vspace{-6mm}
	}
	\maketitle
	\thispagestyle{empty}
	\vspace{-20mm}
	\begin{abstract}
		Semantic communication (SemCom) is emerging as a key technology for future sixth-generation (6G) systems. 
		{Unlike traditional {bit-level} communication (BitCom), SemCom directly optimizes performance at the semantic level, 
        {leading to}  superior {communication} efficiency.} 
		Nevertheless, the task-oriented nature of SemCom 
        {renders }it challenging to completely replace BitCom. 
		Consequently, 
		it is desired to consider a semantic-bit coexisting communication system, where 
		{a base station (BS) serves SemCom users (sem-users) and BitCom users (bit-users) simultaneously.} 
        Such a system faces severe and heterogeneous inter-user interference. 
		In this context, this paper provides a new semantic-bit coexisting communication framework and proposes a spatial beamforming scheme to accommodate both types of users. 
        Specifically, we consider maximizing the semantic rate for semantic users {while ensuring} the quality-of-service (QoS) requirements for bit-users.  
		Due to the intractability of obtaining {the exact closed-form} expression of the semantic rate, a data driven method is first applied to attain an approximated expression via data fitting. 
		With the resulting complex transcendental function, 
        majorization minimization (MM)
        is adopted to convert the original {formulated} problem into a multiple-ratio problem,  
		which allows fractional programming (FP) to be used to further transform the problem into an inhomogeneous quadratically constrained quadratic programs (QCQP) problem. 
		Solving the problem leads to a semi-closed form solution with undetermined Lagrangian factors that can be updated by a fixed point algorithm. 
\rv{This method is referred to as the MM-FP algorithm. Additionally, inspired by the semi-closed form solution, we also propose a low-complexity version of the MM-FP algorithm, called the low-complexity MM-FP (LP-MM-FP), which alleviates the need for iterative optimization of beamforming vectors. Extensive simulation results demonstrate that the proposed MM-FP algorithm outperforms conventional beamforming algorithms such as zero-forcing (ZF), maximum ratio transmission (MRT), and weighted minimum mean-square error (WMMSE). Moreover, the proposed LP-MMFP algorithm achieves comparable performance with the WMMSE algorithm but with lower computational complexity. 
}	\end{abstract}
	
	\vspace{-2mm}
	
	\vspace{-3mm}
	\section{Introduction}
    \vspace{-1mm}
 	\rv{According to Shannon and Weaver \cite{shannon1949mathematical}, communication could be classified into three levels: technical level, which concerns {\itshape{how accurately the symbols of communication are transmitted}};  
	 semantic level, which concerns {\itshape{how precisely the transmitted symbols convey the desired meaning}};  
	 and efficiency level, which concerns {\itshape{how effectively the received meaning affect behavior in the desired way}}. 
	 In the past forty years, researchers mainly focused on the first level design, driving the evolution of mobile communication systems from the first generation (1G) to the fifth generation (5G).} 
	The transmission rate has been significantly improved and the system capacity is gradually approaching the Shannon limit. However, the rapid growth of communication demand in modern society shows no signs of stopping. 
	Specifically, 
	{the upcoming sixth generation (6G) is expected to achieve transmission rates that are ten times faster than those of 5G \cite{jiang2021road}, which will enable the support of numerous new applications}  
	including virtual and augmented reality (VR/AR), smart factories, {intelligent transportation systems}, etc \cite{saad2019vision}. 
	This thus prompts an active research area that rethinks the communication systems at the semantic even effectiveness level. 
	\vspace{-4mm}
	 \subsection{Semantic Communication}
     \vspace{-1mm}
	
	\rv{To build communication systems at the semantic level, semantic communication (SemCom) that mines the semantic information from the source, has emerged as one of the most popular candidate technologies in 6G. SemCom shifts the research focus from compression and transmission of digital bit information to the representation and delivery of semantics, driven by knowledge and logic \cite{chaccour2024less}. 
	The authors in \cite{carnap1952outline} first explored the definition of semantic information, which is based on the logical probability over language content. Building on this definition, the authors in \cite{bao2011towards} further proposed a general transmission paradigm that utilizes the shared knowledge base for SemCom. Then, a semantic communication framework was proposed in \cite{guler2018semantic} to minimize the end-to-end average semantic error. Despite these advancements, SemCom is still in its early stage due to the challenges associated with extracting semantics across common data modalities. }

	\rv{Recently, artificial intelligence (AI) has shown its significant potential in semantic representation and reconstruction. For semantics extraction, the authors in \cite{zhang2023optimization} considered using neural networks to extract the knowledge graph behind images, thereby enabling the effective delivery of semantic information by accurately transmitting the knowledge graph. Given the computation cost of semantic extraction, the authors in \cite{zhao2024joint} further explored joint computation and communication optimization for knowledge graph transmission.} 	
	\rv{For semantics reconstruction, the concept of deep joint source and channel coding (DeepJSCC) has emerged.}
	Compared with the traditional {bit-level} digital communication (BitCom) framework that adopts separate source and channel coding (SSCC) for minimizing bit/symbol error rate, DeepJSCC-based SemCom embraces joint source and channel coding (JSCC) through neural network, which enables {the extraction of} semantic information and {demonstrates} a better transmission efficiency compared with BitCom \cite{xie2021deep}. 
	The authors in \cite{bourtsoulatze2019deep} proposed to use neural network to achieve JSCC for image recovery, and optimized the system performance through end-to-end learning under the criteria of mean square error (MSE). 
    {Then, the authors in \cite{xie2021deep} incorporated transformer and proposed DeepSC, which is shown to outperform 
    BitCom, especially in the low signal-to-noise (SNR) regime.}
	Based on these {pioneering} works \cite{bourtsoulatze2019deep,xie2021deep}, SemCom has then been extensively studied under different data modalities, including image \cite{dai2022nonlinear,gao2023adaptive}, text \cite{jiang2022deep}, speech \cite{weng2021semantic}, video \cite{wang2022wireless}, and multimodal data \cite{zhang2022unified}. 
	{Despite the potential performance gain of SemCom,
    critical concerns about its practical deployment remain.} 
\rv{For example, 
	early JSCC based SemCom systems employ analog  symbol transmission \cite{bourtsoulatze2019deep,xie2021deep,dai2022nonlinear}, 
	while it has been verified that digital transmission is more reliable and secure, 
{as well as cost-effective in hardware {implementation}.}}
    This prompts the development of digital SemCom by designing 
    codebooks \cite{fu2023vector} {and} quantization methods \cite{tung2022deepjscc,bo2022learning} for semantic information. 
	Besides, the current SemCom heavily {relies on} neural networks, which are prone to overfitting the training data collected under certain limited scenarios and thus lack of generalization capability to deal with 
	{
    the challenges brought by the dynamic wireless environment.} 
	Prompted by this, authors in \cite{xu2021wireless} proposed an attention-based JSCC scheme that uses channel-wise soft attention to scale features according to SNR conditions, which enables it {applicable to} scenarios with a broad range of SNRs through a single model. 
	{Then, given} the multiple antenna cases, a channel-adaptive JSCC scheme that exploits the channel state information (CSI) and SNR through attention mechanism was further proposed in \cite{wu2022channel}. 
	\vspace{-4mm}
	\subsection{Motivations}
	\vspace{-1mm}
	Although SemCom has shown great potential for 6G, there is a critical issue {that requires further} investigation: \emph{Can SemCom {completely} replace BitCom?} 
    We believe the answer is no. 
    {This is because the task-oriented nature of SemCom implies that it needs to be tailored for each specific task, which {renders it} not suitable for {generic} transmission tasks.} 
    As a result, we envision that future 6G network will see the co-existence of SemCom and BitCom,  
    yielding the semantic-bit coexisting system that
    {supports both SemCom users (sem-users) and BitCom users (bit-users).} 
    In the coexisting system, due to the {diverse} performance objectives, existing transmission schemes for BitCom {can} no longer provide satisfactory services for the sem-users, and thus need to be redesigned. In response to this, we investigate the beamforming design for the coexisting multi-user multiple-input single-output (MU-MISO) system, and try to shed {lights} on  how to adapt the current transmission algorithms in BitCom to the coexisting system. 
	\vspace{-4.5mm}
	\subsection{Related works}
	\vspace{-1mm} 
{	The study of multiuser SemCom has received a lot of attention in recent years,  
	{which mainly lies in two directions:} resource allocation and interference management. In terms of resource allocation, a semantic-aware channel assignment mechanism was proposed in \cite{yan2022resource}, and an optimal semantic-oriented resource block allocation method was put forward in \cite{liu2022adaptable} subsequently. The main idea of these two works is adjusting {communication resources} for {boosting the transmission of semantic information}. 
    On the other hand, multiuser usually accompanies with interference, which can cause semantic noise that significantly degrades the performance \cite{hu2023robust}. To {mitigate the interference}, several methods have been proposed. For instance, the authors in \cite{zhu2022semantics} proposed to jointly optimize the codebook and the decoder, as such the user interference could be minimized. The authors in \cite{wu2023fusion} further proposed to dynamically fuse the semantic features to a joint latent representation and adjust the weights of different user semantic features to combat fading channels. In addition to the interference from other sem-users, 
    the interference from bit-users {needs to be appropriately mitigated {as well}.} 
    Given this, {the coexistence of sem-users and bit-users was considered in} the non-orthogonal multiple access (NOMA) system \cite{mu2022heterogeneous,li2023non}, where bit-users and sem-users are viewed as primary users and secondary users, respectively. The interference issue was addressed through successive interference cancellation (SIC). 
	{However, in the case of  MU-MISO that enables multi-users through spatial multiplexing,  
    {{the coordination of} the two types of users {with diverse} transmission objectives remains largely unexplored.}}
}	

	\vspace{-0.5mm}
    {
	\rv{Beamforming is a key technique in MU-MISO systems and has been commonly-used for interference mitigation \cite{shi2011iteratively,bjornson2014optimal,xia2019deep,kim2020deep,hu2020iterative,pellaco2021matrix,hu2022irs}.} 
	Several {linear design} methods have been proposed to tackle the beamforming problem in MU-MISO systems. 
    Zero forcing (ZF) and maximum ratio transmission (MRT) algorithm are two simple but effective beamforming algorithms. The former minimizes the user interference, and the latter maximizes the signal gain at the destination user. 
    Besides, a well-known iterative algorithm is the weighted minimum mean-square error (WMMSE) algorithm \cite{shi2011iteratively}, 
    which achieves high performance by 
    first transforming the original problem into an MMSE problem and then updating the variables in an alternative manner. 
	Given high complexity of the WMMSE algorithm and {limited} performance of the ZF and the MRT algorithm, researchers resort to deep learning for developing beamforming scheme with both low complexity and high performance. 
	With the optimal solution structure revealed in \cite{bjornson2014optimal}, the data driven method that learns the undetermined parameters in the solution structure was proposed in \cite{xia2019deep}, and further extended in \cite{kim2020deep}. Besides, the authors in \cite{pellaco2020deep} proposed to use deep unfolding of the WMMSE algorithm for MU-MISO downlink precoding, which constructs the iteration process in neural networks. {Variants} of the deep unfolding-based methods have been investigated in \cite{hu2020iterative}. 
	{The aforementioned schemes aim to maximize {the data rate for BitCom}. However, recent research has {revealed} that the semantic rate in SemCom has a distinct mapping from SNR to performance \cite{yan2022resource,mu2022heterogeneous}.  As a result, existing methods may not be suitable for the semantic-bit coexisting system, and a new beamforming scheme that takes into account the different transmission objectives  is urgently needed.}
	}
	\vspace{-5mm}
	\subsection{Contribution and Organization}
	\vspace{-2mm}
	In this paper, we investigate the transmission design for the semantic-bit coexisting paradigm in the multiple-antenna communication system. 
	Specifically, we consider sem-users with the task of image transmission  
    {and propose an adaptive JSCC autoencoder for semantic information extraction and recovery.} 
    \rv{Recognizing the primary challenge lies in dealing with an intractable semantic rate function, we first perform data regression to model the semantic rate, yielding a complex transcendental function. Then a beamforming problem that optimizes the performance of sem-users under the quality-of-service (QoS) constraints of bit-users is formulated and solved.} 
    The contributions of this paper are summarized as follows:  
	\begin{figure*}[t] 
		\centering
		\includegraphics[width=0.8\linewidth]{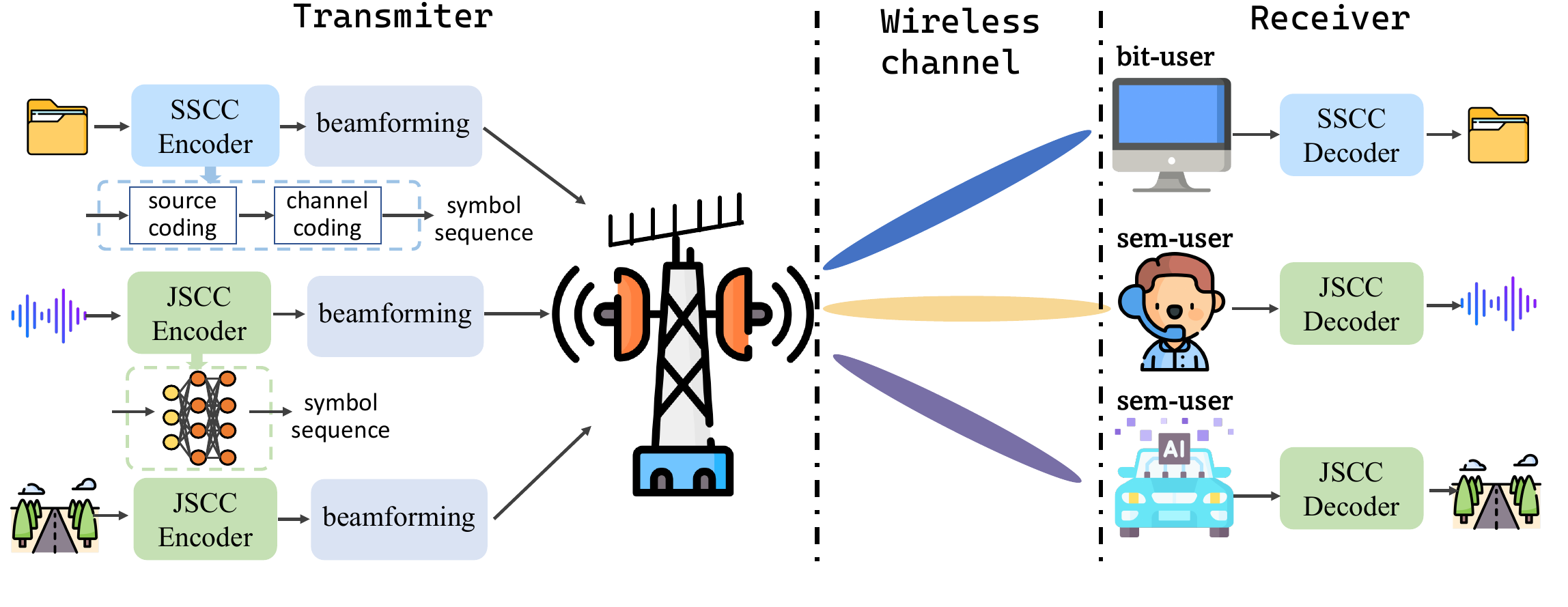}
		\caption{Semantic-bit coexisting communication system framework}
		\vspace{-7.5mm}
		\label{Fig:system framework}
		\end{figure*}
    \begin{itemize}
		\item{Targeting the task of image transmission, we propose an {effective} JSCC scheme that features a dynamic depth of downsampling, which is realized through the ``early exit'' mechanism \cite{zhang2022unified} and the proposed {module-by-module} training scheme. On this basis, we further conduct semantic rate approximation on the ImageNet dataset and build the mapping from the {depth} of downsampling and SNR to semantic rate through data regression. }
		\item{
			We propose a beamforming design scheme for the semantic-bit coexisting system. 
			Specifically, we tackle the primary challenge posed by the transcendental semantic rate function. 
			\rv{By employing majorization-minimization (MM) and introducing a novel surrogate function, the original objective is transformed into a multiple-ratio form, which is further converted to an inhomogeneous quadratically constrained quadratic programs (QCQP) problem by fractional programming.} The semi-closed form solution for the resulting QCQP problem is derived, and the original problem is solved in an alternative manner. Additionally, the alternative algorithm has inspired a low-complexity beamforming method to address the complexity concern.
		}
		\item{
			Both theoretical analysis and numerical simulations are presented to validate the effectiveness of the proposed beamforming scheme in semantic-bit coexisting communication systems. 
		}	
	\end{itemize}
	The rest of this paper is organized as follows. Section \ref{sec: system model} introduces the semantic-bit coexisting system model. 
	Section \ref{sec: JSCC design and semantic rate approximation} presents the proposed JSCC design, the approximation of semantic rate, and the problem formulation. 
	{The optimization problem is solved in} Section \ref{sec: beamforming design}. 
	Then, extensive simulation results are given in Section \ref{sec: experiments}, followed by the concluding remarks in Section \ref{sec: conclusion}.  
	\vspace{-3mm}
    \section{System Model}\label{sec: system model}
    In this section, we first present the semantic-bit coexisting system and the transmission protocol,  
    {based on which} the performance metrics of sem-users and bit-users are analyzed, respectively. 
	\vspace{-5mm}
	\subsection{Semantic-bit Coexisting Communication Framework}\label{subsec: system model of overall framework}


	\rv{We consider a single-cell downlink MU-MISO system shown in Fig. \ref{Fig:system framework}.} The base station (BS) is equipped with $N_t$ transmit antennas, while the users have a single antenna each. The users are divided into two groups, namely bit-users with BitCom and sem-users with SemCom. We denote the bit-users set as $\mathcal{B}=\{b_1,b_2,...,b_B\}$ and the sem-users set as $\mathcal{T}=\{t_1,t_2,...,t_T\}$, with $B$ and $T$ being the numbers of bit-users and sem-users, respectively. 
	The transmit signal vector at the BS, denoted by $\mathbf{x}\in \mathbb{C}^{N_t \times 1}$, is given by 
	\vspace{-3mm}
	\begin{align}\label{eq:transmission signal}
		\mathbf{x} = \sum_{i=1}^B \mathbf{v}_{b_i}{s}_{b_i} + \sum_{j=1}^T\mathbf{v}_{t_j}s_{t_j},
	\end{align}
	where $\mathbf{v}_{b_i}\in \mathbb{C}^{N_t \times 1}$ and $\mathbf{v}_{t_j}\in \mathbb{C}^{N_t \times 1}$ denote the beamforming vector of the $i$-th bit-user and the $j$-th sem-user, respectively. 
    Furthermore, we assume that $s_{b_i}$ and $s_{t_j}$ are zero mean and $\mathbb{E}[s_{b_i}s_{b_i}^H] =\mathbb{E}[s_{t_j}s_{t_j}^H]=1$, and the symbols desired for different users are independent from each other.  

	Then, the received signal $y_{b_k}\in \mathbb{C}$ at bit-user $b_k$ can be expressed as  
	\vspace{-5mm}
	\begin{multline}
		y_{b_k} = \mathbf{h}_{b_k}^H\mathbf{v}_{b_k}s_{b_k} + \sum_{i=1,i\neq k}^{B}\mathbf{h}_{b_k}^H\mathbf{v}_{b_i}s_{b_i} \\+ \sum_{j=1}^{T}\mathbf{h}_{b_k}^H\mathbf{v}_{t_j}s_{t_j} + {n}_{b_k}, \forall b_k \in \mathcal{B}, 
    \end{multline} 
{where $\mathbf{h}_{b_k}\in \mathbb{C}^{N_t \times 1}$ denotes the MISO channel from the BS to user $b_k$ and ${n}_{b_k}$ represents the additive noise which is modeled as a circularly symmetric complex Gaussian random variable following the distribution $\mathcal{CN}\big(\mathbf{0},\sigma_{b_k}^2\big)$, with  $\sigma_{b_k}^2$ being the average noise power. 
}

	Similarly, let $\mathbf{h}_{t_k}\in \mathbb{C}^{N_t \times 1}$ denote the channel from the BS to the sem-user $t_k$, the received signal $y_{t_k}$ at sem-user $t_k$ is given by 
	\vspace{-5mm}
	\begin{multline}
		y_{t_k} = \mathbf{h}_{t_k}^H\mathbf{v}_{t_k}s_{t_k} + \sum_{i=1}^{B}\mathbf{h}_{t_k}^H\mathbf{v}_{b_i}s_{b_i} \\+ \sum_{j=1,j\neq k}^{T}\mathbf{h}_{t_k}^H\mathbf{v}_{t_j}s_{t_j} + {n}_{t_k}, \forall t_k \in \mathcal{T},   
	\end{multline} 
	where ${n}_{t_k}$ is the additive white Gaussian noise with distribution $\mathcal{CN}(0,\sigma_{t_k}^2)$. Notice that {$s_{t_j}$ denotes the symbol stream that contains symbols in the latent representation, which is the output of the JSCC encoder. The symbols within the latent representation are transmitted sequentially.}

	\begin{figure}[h] 
		\centering
		\includegraphics[width=1\linewidth]{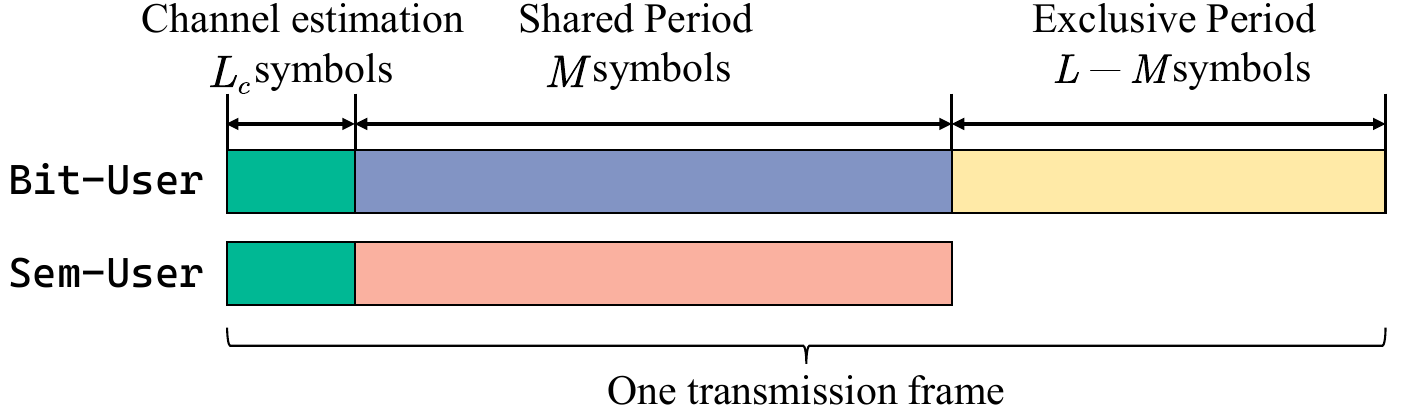}
        \caption{{Transmission protocol for bit-users and sem-users}}
		\vspace{-5mm}
		\label{Fig:frame description}
		\end{figure}
	\subsection{Bit-level Communication}\label{subsec: bit}
	We adopt a transmission frame structure consisting of 
    $L+L_c$ symbol intervals, as shown in Fig. 2. We assume slow fading channel, which means that the channel does not  change within a frame and independently fades across different frames. In this vein, the first $L_c$ symbols are utilized for channel estimation and the remaining $L$ symbols for data transmission. 
	\rv{With the estimated CSI, the BS is able to conduct beamforming.} 
\rv{	It is worth noting that for sem-users, the goal of transmission is to convey the latent representation from the transmitter to the receiver. 
Based on Fig. 10 in \cite{zhang2023predictive} and our own observation, the performance for sem-users does not improve significantly beyond a certain threshold of the number of transmitted symbols. Drawing inspiration from semi-NOMA principle outlined in \cite{mu2023exploiting}, we assume that sem-users complete transmission of the latent representation within $M$ symbol intervals ($M < L$), while bit-users utilize all $L$ symbol intervals for data transmission.
}	
	In this context, the total data transmission component of length $L$ symbol intervals is further divided into two parts, as shown in Fig. \ref{Fig:frame description}. At the shared period of length $M$ symbol intervals, the BS simultaneously serves all users. Both bit-users and sem-users will be interfered by each other.  
    {The exclusive period of length $L-M$ symbol intervals is dedicated to data transmission for bit-users, i.e., $s_{t_k}=0, \forall t_k\in \mathcal{T}$.
	\footnote{\rv{Note that while we primarily focus on the scenario where the number of transmission symbols used by bit-users exceeds that of sem-users, it should be emphasized that our framework and proposed method can be readily extended to situations where the number of transmission symbols used by sem-users is greater than that of bit-users. This extension can be achieved by 
	revising the overall bit rate of the bit-users $b_i$ in (\ref{eq: rate performance}) to $R_{b_i}=\frac{M}{L}\log_2 (1+\gamma_{b_i,1})$. 
}	}}
	As digital transmission is employed at the bit-user,  the achievable bit rate (bits/s/Hz) {during} the shared period is given by \cite{shannon1949mathematical}
	\begin{align}\label{eq: rate performance of the first part}
		R_{{b_i,1}} = \log_2 \left(1+{\gamma}_{b_i,1}\right), 
	\end{align}
	where $\gamma_{b_i,1}=\frac{|\mathbf{h}_{b_i}^H\mathbf{v}_{b_i}|^2}{\sum_{j\in \{\mathcal{B},\mathcal{T}\}/b_i}|\mathbf{h}_{b_i}^H\mathbf{v}_j|^2+\sigma_{b_i}^2}$ denotes the signal-to-interference-plus-noise ratio (SINR) of bit-user $b_i$ during the shared period. 

	Then, at the exclusive period, the BS only serves the bit-users, 
    and the corresponding achievable bit rate is given by 
	\begin{align}\label{eq: rate performance of the second part}
		R_{{b_i,2}} = \log_2 \left(1+{\gamma}_{b_i,2}\right), 
	\end{align}
	where $\gamma_{b_i,2}=\frac{|\mathbf{h}_{b_i}^H\mathbf{v}_{b_i}|^2}{\sum_{j\in \{\mathcal{B}\}/b_i}|\mathbf{h}_{b_i}^H\mathbf{v}_j|^2+\sigma_{b_i}^2}$. 

	As a result, the overall normalized bit rate of the bit-user $b_i$ in a frame is defined as below. 
	\begin{align}\label{eq: rate performance}
		R_{{b_i}} = \frac{M}{L}\log_2 \left(1+{\gamma}_{b_i,1}\right) + \frac{L-M}{L}\log_2 \left(1+{\gamma}_{b_i,2}\right).  
	\end{align}
	\vspace{-10mm}
	\subsection{Semantic Communication}

	In semantic communication, the semantic rate no longer focuses on the {symbol error rate}, but on the quality of task completion. 
    Fundamentally, the performance of semantic communication hinges on the effectiveness of the JSCC model and the wireless noise intensity. 
    {In this sense}, 
	the semantic rate can be generally expressed as 
    \vspace{-2mm}
	\begin{align}\label{eq:semantic rate general expression}
		S_{t_i} = \epsilon(\boldsymbol{\Theta},\gamma_{t_i}), 
	\end{align}
	where $\boldsymbol{\Theta}$ denotes the semantic model composed of deep neural networks (DNNs) that determines $M$ and the specific {method for the extraction} of semantic information, and $\gamma_{t_i}=\frac{|\mathbf{h}_{t_i}^H\mathbf{v}_{t_i}|^2}{\sum_{j \in \{\mathcal{B,T}\}/t_i }|\mathbf{h}_{t_i}^H \mathbf{v}_j|^2+\sigma_{t_i}^2}$ is the SINR of the sem-user $t_i$ at the shared period. 
	In the context of image transmission scenario, $\epsilon(\boldsymbol{\Theta},\gamma_{t_i})$ is evaluated under the widely-adopted performance metric called structural similarity index measure (SSIM). 

	As mentioned, the overall semantic rate is determined by the adopted JSCC model and the transmission environment.  
	The former represents the semantic compression and exploitation ability of the semantic communication, while the later determines the level of noise disturbation. 
{However, semantic communication} highly relies on neural networks for semantic extraction and recovery, the black-box nature of which hinders the theoretical analysis, making $\epsilon(\boldsymbol{\Theta},\gamma_{t_i})$ unable to be acquired precisely. 
	{A commonly adopted} method for tackling this problem is data regression \cite{mu2022heterogeneous,mu2023exploiting,yan2022resource}, which obtains the mapping from $\boldsymbol{\Theta}$ and $\gamma_{t_i}$ to $S_{t_i}$ through sufficient experimental instances and curve fitting,  
	{which will be further elaborated in the subsequent section.}
	\vspace{-2mm}
    \section{JSCC Design and Semantic Rate Approximation}\label{sec: JSCC design and semantic rate approximation}
	In this section, we will elaborate on the design of semantic communication in detail. 
	Considering the task of image transmission,
	we first present the proposed design of the JSCC model. 
	Then, we conduct a series of experiments to evaluate the performance in different system settings. 
	Building on this, we approximate the semantic rate with data regression. 
    Finally, the problem that jointly optimizes the beamforming vectors and the  downsampling depth is formulated. 
	\vspace{-3mm}
	\subsection{JSCC Network Characterization}
	\begin{figure}[h] 
		\centering
		\includegraphics[width=1\linewidth]{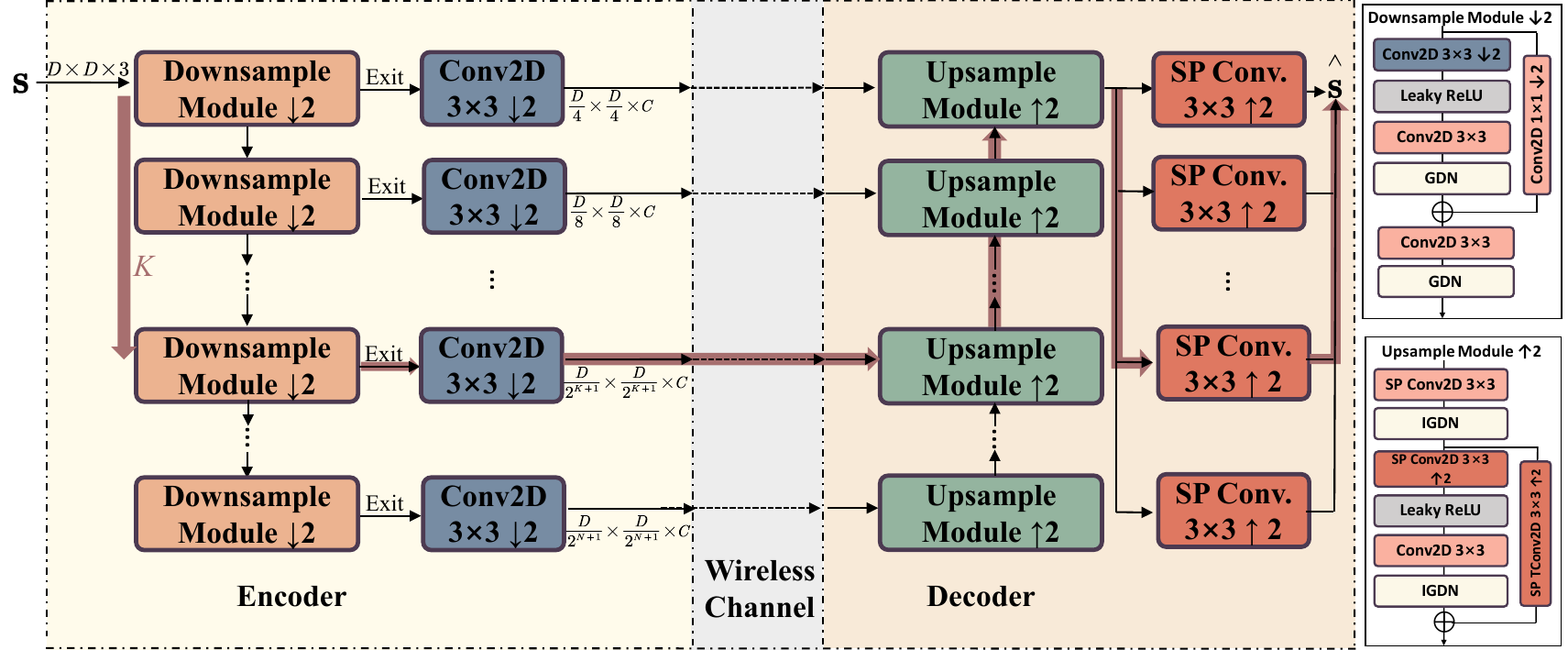}
		\vspace{-1mm}
		\caption{{JSCC network for image transmission with multi exit mechanism}}\label{fig: JSCC network}
		\vspace{-5mm}
		\label{Fig:JSCC network}
		\end{figure}
	The proposed JSCC network for image transmission is shown in Fig. \ref{fig: JSCC network}. 
	At the encoder part, we consider compressing the original image $\mathbf{s}$ through multiple downsample modules $\mathcal{D}=\{\mathscr{D}_n\}_{n=1}^N$,  
	each of which comprises a residual block \cite{cheng2019deep}, followed by a convolution layer. 
	The number of filters in all the convolution layers is set to $C$. 
	After each downsample module, the image size is reduced by half, and the number of channels is fixed to $C$. For the upsample module $\mathcal{U}=\{\mathscr{U}_n\}_{n=1}^N$ at the decoder part, the reverse process is conducted. 
	Without loss of generality, 
	we consider the image having a square size, that is, $\mathbf{s}\in \mathbb{R}^{I\times I\times 3}$, with $I$ being the image size, 
	and ``3'' the number of channels, 
    respectively.

\rv{Additionally, in typical 5G and beyond communication systems, due to the rapid growth in the number of users and data {volume}, 
careful resource allocation is needed at the BS side. 
}	As a result, the available communication resources for users vary considerably in time and space,  which poses new requirements for semantic communication. Specifically, the JSCC model should be able to dynamically adjust the number of the transmission symbols. 
    {To this end}, we propose a multi-exit mechanism, as illustrated in Fig. 3.  
	With this mechanism, the decoder can exit early rather than pass all the downsample modules. Consequently, the {size} of the latent representation can be {adjusted} by selecting the number of passed downsample module. Let $K$ be the number of passed downsample module, then the number of required {transmission} symbols $M_K$ is given by 	
	\vspace{-2mm}
    \begin{align}\label{eq: calculation of M}
		M_K=C\big(\frac{I}{2^{K+1}}\big)^2=\frac{CI^2}{4^{K+1}}. 
	\end{align}
	\vspace{-5mm}
	\subsection{Semantic Rate Approximation}\label{subsec:semantic rate approximation}
	Before deployment, 
    {training is required to obtain the JSCC neural network $\boldsymbol{\Theta}$}. 
    With  the multi-exit mechanism, it is desired that the downsample and upsample modules in $\boldsymbol{\Theta}$ can work independently, and also be incorporated into deeper models (i.e., $\boldsymbol{\Theta}$ with a larger $K$). 
	To this end, we propose a module-by-module training algorithm. As shown in Algorithm \ref{algo:algorithm for module-by-module training}, the modules are trained sequentially, with only the weight parameters in the current layer modules being updated during the $K$-th round of training, while the upper layer modules (i.e., $\{\mathscr{D}_n\}_{n=1}^{K-1}$, $\{\mathscr{U}_n\}_{n=1}^{K-1}$) are frozen\footnote{``Frozen'' means that the parameters of the module will not change anymore.}. 
    {Let $\boldsymbol{\Theta}_K$ be the JSCC model with a specific downsampling depth $K$. The semantic rate defined in (\ref{eq:semantic rate general expression}) is given by $\epsilon(\boldsymbol{\Theta}_K, \gamma)$, where $\gamma$ denotes the SNR. 
    For simplicity, we use the notation $\epsilon(K, \gamma)$ for $\epsilon(\boldsymbol{\Theta}_K, \gamma)$ in the rest of this paper, since $\boldsymbol{\Theta}_K$ is uniquely determined by $K$ when the downsampling and upsampling modules are specified. 
    }
	\begin{figure}[!]
		\label{alg:LSB}
		\begin{algorithm}[H]
		  \caption{Algorithm for Module-by-module Training}\label{algo:algorithm for module-by-module training}
		  \begin{algorithmic}[1]
			\renewcommand{\algorithmicrequire}{ \textbf{Initialize}}
			\REQUIRE all the modules including $\{\mathscr{D}_M\}_{M=1}^{N}$, $\{\mathscr{U}_M\}_{M=1}^{N}$. 
			\FOR {$K=1, 2, ..., N$}
			\STATE $\Theta=\{\{\mathscr{D}_M\}_{M=1}^{K},\{\mathscr{U}_M\}_{M=1}^K\}$. 
			\WHILE{$\Theta$ not converged}
			\STATE Sample a minibatch of data from training dataset
			\STATE Obtain the gradient $\nabla \mathscr{D}_K$ and $\nabla \mathscr{U}_K$ through loss calculation and backpropagation. 
			\STATE Update $\mathscr{D}_K$ and $\mathscr{U}_K$ using 
			$\mathscr{D}_K = \mathscr{D}_K - \eta \nabla \mathscr{D}_K$, $\mathscr{U}_K = \mathscr{U}_K - \eta \nabla \mathscr{U}_K$. 
			\ENDWHILE 
			\STATE Frozen $\mathscr{D}_K$ and $\mathscr{U}_K$. 
			\ENDFOR
		  \end{algorithmic}
		\end{algorithm}
		\vspace{-10mm}
	\end{figure}

\begin{figure}[h] 
		\centering
		\includegraphics[width=1\linewidth]{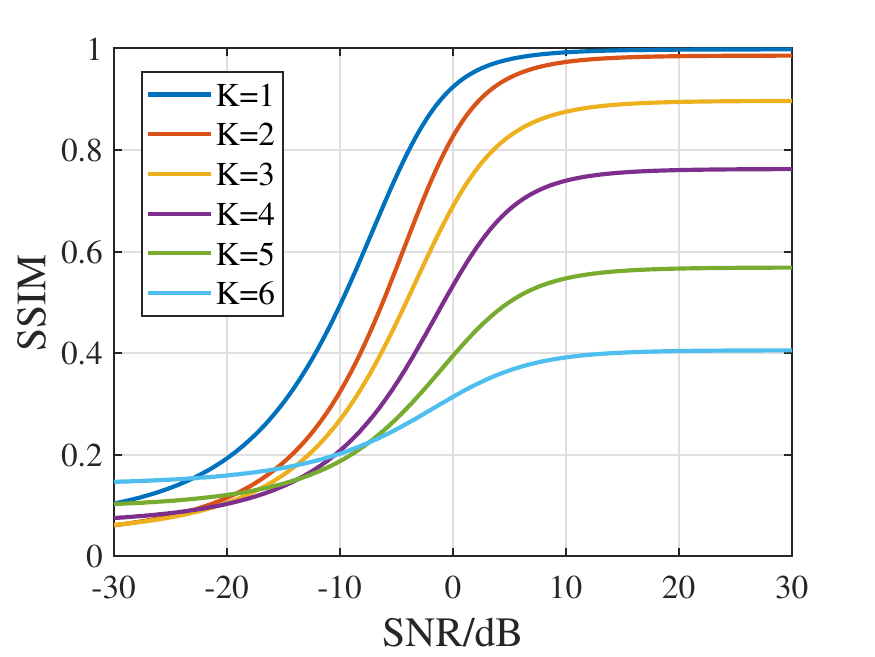}
        \vspace{-1mm}
		\caption{Performance evaluation under different $K$ and SNR settings}
		\vspace{-8mm}
		\label{Fig:performance on imagent}
		\end{figure}

	We train $\boldsymbol{\Theta}$ on the ImageNet dataset \cite{deng2009imagenet}, a large-scale image dataset containing over 14 million images, which serves as a standard benchmark for various computer vision tasks. 
	\rv{We consider the additive white Gaussian noise (AWGN) channel}\footnote{Note that {we assume AWGN channel for simplicity, such that} 
    the JSCC model can be trained on the BS side, and the training overhead can thus be significantly reduced. Nevertheless, as we discussed in the previous work \cite{zhang2023wireless}, the model trained under AWGN cases can be directly applied to the MISO cases with minor modification. This is because the final received signal can be transformed into an equivalent AWGN form when recovery precoding is adopted at the receiver.}, where the SNR is fixed to 10 dB in the training process. Moreover, \emph{mean squared error} (MSE) criteria is adopted as the loss function, {and the Adam optimizer with the initiate learning rate of $10^{-4}$ is adopted}. 
    {After sufficient training, we evaluate the performance $\epsilon(K,\gamma)$ on the {validation} dataset with different $K$ and SNR settings, under SSIM.} The evaluation results are depicted in Fig. \ref{Fig:performance on imagent}. 
\rv{    It can be seen that with an increasing $K$, i.e., {more stringent compression of the original image, the performance floor when ${\rm SNR}\rightarrow \infty$ decreases monotonically.}
}    
\rv{Besides, for each $K$, $\epsilon(K,\gamma)$ follows an $S$ shape with respect to $\gamma$ in dB, which is also revealed in \cite{mu2022heterogeneous,mu2022exploiting,getu2024performance} under the text transmission task with the DeepSC model \cite{xie2021deep}. Therefore, similar to \cite{mu2022heterogeneous}, the generalized logistic function could be utilized to well approximate $\epsilon(K,\gamma)$, as follows. 
}	
\rv{\begin{align}\label{eq: the gt semantic rate function}
		\epsilon(K,\gamma) \approx \tilde{\epsilon}(K,\gamma)&= a_K + \frac{d_K}{c_K+{\rm exp}(-10\alpha_K\log_{10}(\gamma))}\notag\\&=a_K + \frac{d_K}{c_K+\gamma^{-e_K}}, 
\end{align}
}	where $a_K$, $c_K$, $d_K$, $e_K$ are parameters determined by $K$, and are obtained through curve fitting, and we have $e_K=\frac{10\alpha_K}{\ln{10}}$.  
	\vspace{-3.5mm}
    \vspace{-1mm}
	\subsection{Problem Formulation}
	As illustrated in Section \ref{subsec: system model of overall framework}, 
	{there are two types of users with different performance metrics in the sematic-bit coexisting communication system}. 
	In this paper, we aim to maximize the semantic rate of sem-users while satisfying the QoS requirements of bit-users, 
	by jointly optimizing the beamforming vectors and the  downsampling depth. 
	The considered optimization problem can be formulated as 
	\vspace{-2mm}
	\begin{subequations}\label{eq:problem formulation}
		\begin{align}
			\mathbf{P1}: \max_{\mathbf{V},K} ~~ &\sum_{i=1}^T \tilde{\epsilon} \left(K,\gamma_{t_i}\right),\\
			{\rm s. t.}~~&
            \frac{M_K}{L}\log_2\left(1+\gamma_{b_i,1}\right)+\frac{L-M_K}{L}\notag\\&\qquad~~~~\cdot\log_2\left(1+\gamma_{b_i,2}\right) \geq \beta_{b_i}, \forall b_i \in \mathcal{B},\label{eq: qos constraints in p1}\\
            &\sum_{i=1}^B\left\|\mathbf{v}_{b_i}\right\|^2 + \sum_{j=1}^T\left\|\mathbf{v}_{t_j}\right\|^2 \leq P_T, \label{eq: power constraints in p1}
			\end{align}
			\end{subequations}
		where {$\gamma_{t_i}=\frac{|\mathbf{h}_{t_i}^H\mathbf{v}_{t_i}|^2}{\sum_{j \in \{\mathcal{B,T}\}/t_i }|\mathbf{h}_{t_i}^H \mathbf{v}_j|^2+\sigma_{t_i}^2}$}, $\gamma_{b_i,1}=\frac{|\mathbf{h}_{b_i}^H\mathbf{v}_{b_i}|^2}{\sum_{j\in \{\mathcal{B},\mathcal{T}\}/b_i}|\mathbf{h}_{b_i}^H\mathbf{v}_j|^2+\sigma_{b_i}^2}$, $\gamma_{b_i,2}=\frac{|\mathbf{h}_{b_i}^H\mathbf{v}_{b_i}|^2}{\sum_{j\in \{\mathcal{B}\}/b_i}|\mathbf{h}_{b_i}^H\mathbf{v}_j|^2+\sigma_{b_i}^2}$, and $\mathbf{V}=[\mathbf{V}_B,\mathbf{V}_T]\in \mathbb{C}^{N_t\times (B+T)}$, with $\mathbf{V}_B=[\mathbf{v}_{b_1},...,\mathbf{v}_{b_B}]\in \mathbb{C}^{N_t\times B}$ and $\mathbf{V}_T=[\mathbf{v}_{t_1},...,\mathbf{v}_{t_T}]\in\mathbb{C}^{N_t\times T}$. $\beta_{b_i}$ denotes the requirements of transmission rate in one frame from bit-user $b_i$. $P_T$ denotes the transmit power budget of the BS. $R_{b_i,1}$ and $R_{b_i,2}$ denote the transmission rate defined in (\ref{eq: rate performance of the first part}) and (\ref{eq: rate performance of the second part}), respectively. 
        {$\tilde{\epsilon}(K,\gamma_{t_i})$ is the approximated semantic rate given by (\ref{eq: the gt semantic rate function}).} 
	\vspace{-1mm}
		\begin{remark}\label{remark: problem formulation discussion}
		\emph{
		As shown in $\mathbf{P1}$, beamforming design for a semantic-bit coexisting system faces some new challenges compared to a BitCom system. 
        {Firstly, the semantic rate $\tilde{\epsilon}(K,\gamma_{t_i})$ admits a completely different form (which is neither convex nor concave) from channel capacity w.r.t. SINR, which renders the existing interference suppression algorithms ineffective.}  
        Additionally, the performance of sem-users also partially depends on the downsampling depth {$K$ that requires careful design}. Unfortunately, there exists a strong coupling between beamforming design and $K$, making the problem even more challenging to solve.  
		}
	\end{remark}
	\vspace{-4.5mm}
		\section{{Joint Optimization of Beamforming and $K$-Configuring for Coexisting System}}\label{sec: beamforming design}
		In this section, we solve the problem $\mathbf{P1}$ for beamforming design and configuring $K$ in semantic-bit coexisting MU-MISO systems. 
	As discussed in Remark \ref{remark: problem formulation discussion}, it is hard to directly solve the joint optimization problem, and we thus consider solving $\mathbf{P1}$ by optimizing $\mathbf{V}$ and $K$ {alternatively}. 

	\vspace{-3mm}
	\subsection{Beamforming Design}
	In this subsection, we optimize the beamforming matrix $\mathbf{V}$ in $\mathbf{P1}$ with a given $K$, and the subproblem is given below. 
	\vspace{-1.5mm}
	\begin{subequations}\label{eq: subproblem for beamforming optimization}
        \begin{align}
		\mathbf{P2}: \max_{\mathbf{V}}~~ &\sum_{i=1}^T a_K+\frac{d_K}{c_K+(\gamma_{t_i})^{-e_K}},\label{eq:opjective function in p2}\\
		{\rm s.t.}~~&(\ref{eq: qos constraints in p1}), (\ref{eq: power constraints in p1}). \notag
		\end{align}
	\end{subequations}
	As shown in $\mathbf{P2}$, the objective function (\ref{eq:opjective function in p2}) is non-convex as (\ref{eq:opjective function in p2}) is a transcendental function of $\mathbf{V}$. 
	Moreover, the fractional expression exists in the QoS constraints. 
	Therefore, $\mathbf{P2}$ is a NP-hard problem, indicating that the optimal solution is intractable. 
	We thus resort to a suboptimal solution. 
	{To this end, the} {problem-solving} process is mainly divided into four steps. 
	Firstly, 
    {we {relax} the power constraint by regulating the noise intensity with ${{\rm Tr}(\mathbf{V}\mathbf{V}^H)}/{P_T}$.} 
{Then, we propose a surrogate function for approximating the objective function.  
	Next, the transforming method proposed in \cite{shen2018fractional} is adopted to transform the multiple-ratio fractional programming (FP) problem into a QCQP problem. 
    Finally, the resulting QCQP problem is solved in a low-complexity manner. }
	\subsubsection{Problem {Transformation}}
{    It can be observed that the beamforming vectors appear in $\mathbf{P}_2$ as 
}    
    the form of SINR in both the objective function (\ref{eq:opjective function in p2}) and the QoS constraints (\ref{eq: qos constraints in p1}). 
	Without loss of optimality, similar to \cite{hu2020iterative,zhao2023rethinking}, the power constraint  can be removed by integrating it to the SINR terms, as follows. 
	\vspace{-2mm}
    \begin{subequations}\label{eq: subproblem for beamforming optimization after remove power constraints}
		\begin{align}
			\mathbf{P3}: \max_{\mathbf{V}} \qquad &\sum_{i=1}^T a_K+\frac{d_K}{c_K+(\Gamma_{t_i})^{-e_K}},\label{eq:objective function of p3}\\
			{\rm s.t.} \qquad&\frac{M_K}{L}\log_2\left(1+\Gamma_{b_i,1}\right)+ \frac{L-M_K}{L}\notag \\ &\qquad\cdot\log_2\left(1+\Gamma_{b_i,2}\right) \geq \beta_{b_i}, \forall b_i \in \mathcal{B}.\label{eq: qos constraints in p3} 
			\end{align}
		\end{subequations}
    \vspace{-10mm}
    where the equivalent SINR terms are given by
    \vspace{8mm}
    \begin{align}
		\Gamma_{t_i}&= \frac{|\mathbf{h}_{t_i}^H\mathbf{v}_{t_i}|^2}{\sum_{j \in \{\mathcal{B,T}\}/t_i }|\mathbf{h}_{t_i}^H \mathbf{v}_j|^2+\frac{{\rm Tr}(\mathbf{V}\mathbf{V}^H)}{P_T}\sigma_{t_i}^2},\label{eq: Gamma ti formulation} \\
        \Gamma_{b_i,1} &=\frac{\left|\mathbf{h}_{b_i}^H\mathbf{v}_{b_i}\right|^2}{\sum_{j
        \in\left\{\mathcal{B},
        \mathcal{T}\right\}/b_i }\left|\mathbf{h}_{b_i}^H \mathbf{v}_{j}\right|^2+\frac{{\rm Tr}(\mathbf{V}\mathbf{V}^H)}{P_T}\sigma_{b_i}^2},\label{eq: Gamma bi1 formulation} \\
        \Gamma_{b_i,2} &= \frac{\left|\mathbf{h}_{b_i}^H\mathbf{v}_{b_i}\right|^2}{\sum_{j
        \in\left\{\mathcal{B}\right\}/b_i }\left|\mathbf{h}_{b_i}^H \mathbf{v}_{j}\right|^2+\frac{{\rm Tr}(\mathbf{V}\mathbf{V}^H)}{P_T}\sigma_{b_i}^2}.     \label{eq: Gamma bi2 formulation}
    \end{align} 

    \begin{figure*}[t]
        \begin{align}
            \zeta_K(\mathbf{V},\Gamma_{t_i}^0) = D(K,\Gamma_{t_i}^0)+E(K,\Gamma_{t_i}^0)\frac{|\mathbf{h}_{t_i}^H\mathbf{v}_{t_i}|^2}{F(K,\Gamma_{t_i}^0)|\mathbf{h}_{t_i}^H\mathbf{v}_{t_i}|^2+G(K,\Gamma_{t_i}^0)(\sum_{j\in \{\mathcal{B,T}\}/t_i }|\mathbf{h}_{t_i}^H\mathbf{v}_j|^2+\frac{{\rm Tr}(\mathbf{V}\mathbf{V}^H)}{P_T}\sigma_{t_i}^2)},\label{eq: objective sort result}
        \end{align}
        \hrule
		\vspace{-6mm}
    \end{figure*}
	\rv{Let $\mathbf{V}^{*}$ and $\mathbf{V}^{**}$ denote the optimal solutions of problems $\mathbf{P2}$ and $\mathbf{P3}$, respectively. 
By observing $\Gamma_{t_i}$, $\Gamma_{b_i,1}$, and $\Gamma_{b_i,2}$ in (\ref{eq: Gamma ti formulation}), (\ref{eq: Gamma bi1 formulation}), and (\ref{eq: Gamma bi2 formulation}), 
it can be inferred that $\alpha \mathbf{V}^{**}$ should also be an optimal solution for problem $\mathbf{P3}$, where $\alpha$ is a scaling factor. 
When $\alpha$ serves as a power normalization scalar, i.e., $\alpha=\sqrt{\frac{{P_T}}{{\rm Tr}(\mathbf{V}^{**}(\mathbf{V}^{**})^H)}}$, $\alpha \mathbf{V}^{**}$ achieves the maximum value of the objective function in problem $\mathbf{P2}$, as $\alpha \mathbf{V}^{**}$ maximizes the objective function of problem $\mathbf{P3}$. 
Moreover, it is straightforward to validate that $\alpha \mathbf{V}^{**}$ also satisfies (\ref{eq: qos constraints in p1}), (\ref{eq: power constraints in p1}). 
Therefore, we can conclude that $\mathbf{V}^{*} = \alpha \mathbf{V}^{**}$. 
More importantly, this allows us to solve problem $\mathbf{P2}$ by first solving problem $\mathbf{P3}$ and then applying power normalization to the solution.
}	
	\subsubsection{Objective Approximation}
	\rv{Observing problem $\mathbf{P3}$, it can be found that the objective function (\ref{eq:objective function of p3}) presents a complex transcendental form that can not be directly tackled. To address it, 
	Observing problem $\mathbf{P3}$, it can be found that the objective function (12a) presents a complex transcendental form that can not be directly tackled. To address it, 
we first employ the MM algorithm \cite{sun2016majorization} to solve the problem $\mathbf{P3}$ in an alternative manner and then use a surrogate function to approximate the objective function.}
    Firstly, the semantic rate in $\mathbf{P3}$ is given by 
	\vspace{-3mm}
	\begin{align}
		\tilde{\epsilon}_K(\Gamma_{t_i}) = a_K + \frac{d_K}{c_K+(\Gamma_{t_i})^{-e_K}}. 
	\end{align}
	\begin{figure}[h] 
		\centering
		\subfigure[$e_K\leq 1$]{
			\includegraphics[width=0.478\linewidth]{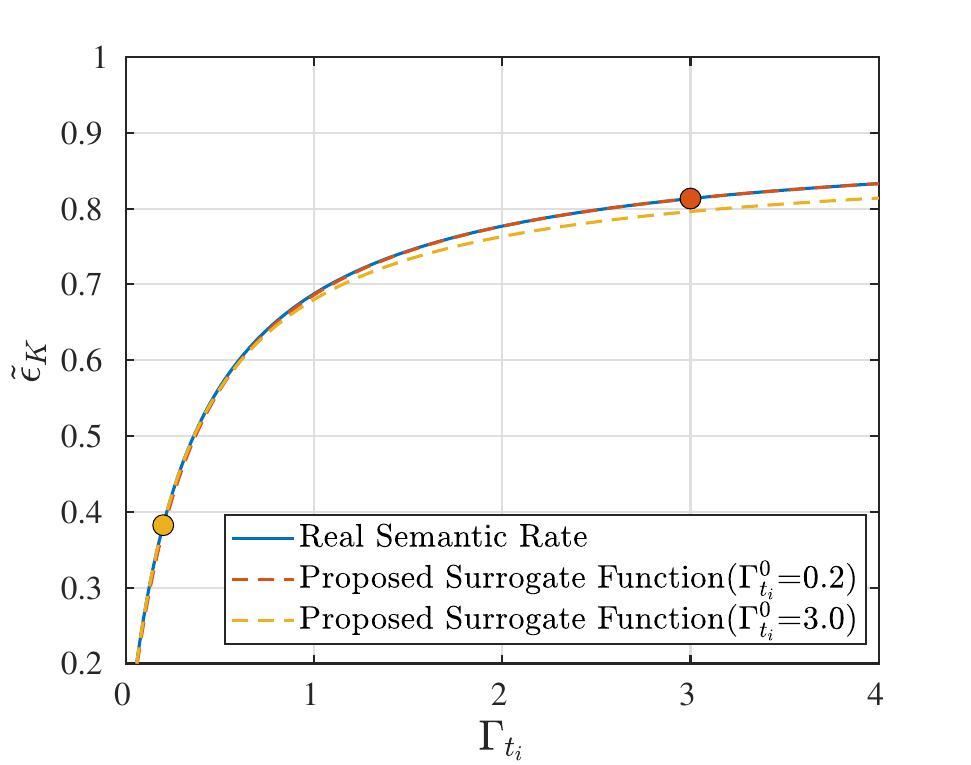}}
		\subfigure[$e_K >1$]{
			\includegraphics[width=0.475\linewidth]{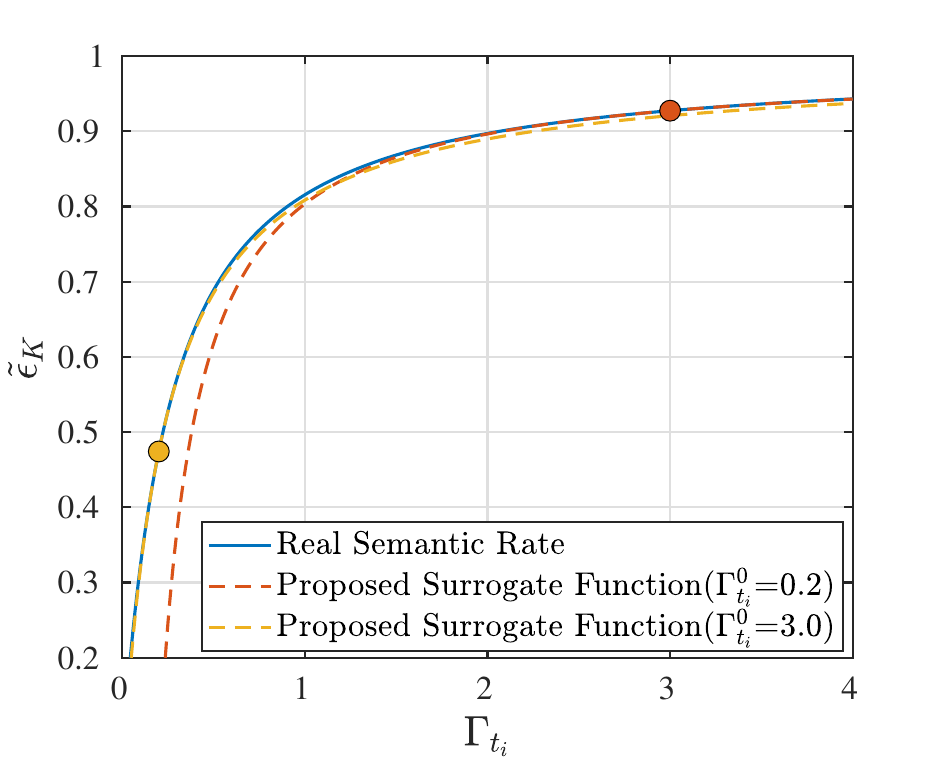}}
		\vspace{1mm}
        \caption{{Objective Approximation with different surrogate functions}}
        \vspace{-3.5mm}
		\label{Fig:objective approximation}
	\end{figure}
	The exemplary functions with different $K$ settings are depicted in Fig. \ref{Fig:objective approximation}. 
    {Note that, our objective is to maximize the semantic rate w.r.t. the precoding matrix $\mathbf{V}$ rather than the SINR term $\{\Gamma_{t_i}\}$. 
	Therefore, it is necessary to find an appropriate surrogate function that can accurately capture the shape of $\epsilon_K(\Gamma_{t_i})$ and also has a simple form for ease of handling during the optimization of $\mathbf{V}$.} 
\rv{    To this end, 
	we propose the following surrogate function to approximate $\tilde{\epsilon}_K(\cdot)$ at a given station point $\Gamma_{t_i}^{0}$. 
}	
\rv{\begin{proposition}\label{prop: lower bound of semantic rate}
	\emph{A lower bound on the semantic rate function $\tilde{\epsilon}_K(\Gamma_{t_i})$ is given by}
	\begin{align}\label{eq: proposition 1 equation}
		\tilde{\epsilon}_K(\Gamma_{t_i})\geq \zeta_K(\Gamma_{t_i},\Gamma_{t_i}^0),
	\end{align}
	\emph{where the equality holds only when $\Gamma_{t_i}=\Gamma_{t_i}^0$.  
	$\zeta_K(\Gamma_{t_i},\Gamma_{t_i}^0)=a_K+\frac{d_K}{c_K+\mathcal{J}(\Gamma_{t_i},{\Gamma_{t_i}^0},e_K)}$, 
	and $\mathcal{J}(\Gamma_{t_i},{\Gamma_{t_i}^0},e_K)$ is given as follows.}
    \begin{small}
		\begin{align}\label{eq:surrogate function}
			\mathcal{J}(\Gamma_{t_i},{\Gamma_{t_i}^0},e_K)=\left\{
			\begin{aligned}
				\frac{e_K\Gamma_{t_i}^0+(1-e_K)\Gamma_{t_i}}{(\Gamma_{t_i}^0)^{e_K}\Gamma_{t_i}},e_K\leq 1, \\
				\frac{1}{(\Gamma_{t_i}^0)^{e_K-1}(\Gamma_{t_i}^0+e_K(\Gamma_{t_i}-\Gamma_{t_i}^0))},e_K>1,
			\end{aligned}
			\right.
		\end{align}
	\end{small}	
\end{proposition}
}	
\begin{proof}
	\rv{To establish Proposition \ref{prop: lower bound of semantic rate}, we only need to prove $(\Gamma_{t_i})^{-e_K}\leq \mathcal{J}(\Gamma_{t_i},{\Gamma_{t_i}^0},e_K)$. }

\rv{For $e_K\leq 1$, $f(x)=x^{e_K}$ is a concave function for $x\geq0$. Let $x=\frac{1}{\Gamma_{t_i}}$,  then we have $\big(\frac{1}{\Gamma_{t_i}}\big)^{e_K}\leq \big(\frac{1}{\Gamma_{t_i}^{0}}\big)^{e_K}+e_K\big(\frac{1}{(\Gamma_{t_i}^0)}\big)^{e_K-1}\big(\frac{1}{\Gamma_{t_i}}-\frac{1}{\Gamma_{t_i}^0}\big)$, the equality holds if and only if $\Gamma_{t_i}=\Gamma_{t_i}^0$. By sorting this result, we can conclude that Proposition \ref{prop: lower bound of semantic rate} holds for $e_K\leq 1$.}

\rv{For $e_K >1$, $f(x)=x^{e_K}$ is a convex function for $x\leq 0$. Let $x=\Gamma_{t_i}$, then we have $(\Gamma_{t_i})^{e_K}\geq (\Gamma_{t_i}^0)^{e_K}+e_K(\Gamma_{t_i}^0)^{e_K-1}(\Gamma_{t_i}-\Gamma_{t_i}^0)$, the equality holds if and only if $\Gamma_{t_i}=\Gamma_{t_i}^0$. By sorting this result, we can conclude that Proposition \ref{prop: lower bound of semantic rate} holds for $e_K> 1$.}

\rv{In a nutshell, Proposition \ref{prop: lower bound of semantic rate} holds for any $e_K>0$, which ends the proof.}
\end{proof}
Prompted by Proposition \ref{prop: lower bound of semantic rate},
we use $\zeta_K(\Gamma_{t_i},\Gamma_{t_i}^0)$ as the surrogate function for $\tilde{\epsilon}_K(\Gamma_{t_i})$. 
    As shown in Fig. \ref{Fig:objective approximation}, the proposed surrogate function captured the original $\tilde{\epsilon}_K(\cdot)$ well. 
	We further take $\Gamma_{t_i}$ in (\ref{eq: Gamma ti formulation}) into $\zeta_K(\Gamma_{t_i},\Gamma_{t_i}^0)$, which yields $\zeta_K(\mathbf{V},\Gamma_{t_i}^0)$ and 
	{leads to the following optimization problem: } 
	\vspace{-2mm}
    \begin{subequations}\label{eq: subproblem for beamforming optimization}
			\begin{align}
				\mathbf{P4}: \max_{\mathbf{V}}  &\sum_{i=1}^T \zeta_K(\mathbf{V},\Gamma_{t_i}^0),\label{eq:approximated objective}\\
				{\rm s.t.}\qquad&(\ref{eq: qos constraints in p3}).\notag
				\end{align}
		\end{subequations}
        It can be found that 
        $\zeta_K(\cdot)$ exhibits a fractional form of $\Gamma_{t_i}$. Therefore, we resort $\zeta_K(\cdot)$ as a function of beamforming vectors, as shown in (\ref{eq: objective sort result}) on the top of this page, 
		where $D(K,\Gamma_{t_i}^0)=a_K$, $E(K,\Gamma_{t_i}^0)=d_K$, $F(K,\Gamma_{t_i}^0)=c_K + (1-e_K)(\Gamma_{t_i}^{0})^{-e_K}$, and $G(K,\Gamma_{t_i}^0)=e_K(\Gamma_{t_i}^{0})^{1-e_K}$ when $e_K \leq 1$; 
		$D(K,\Gamma_{t_i}^0)=a_K+\frac{d_K(1-e_K)(\Gamma_{t_i}^0)^{e_K}}{c_K(1-e_K)(\Gamma_{t_i}^0)^{e_K}+1}$, $E(K,\Gamma_{t_i}^0)=\frac{d_Ke_K(\Gamma_{t_i}^0)^{e_K-1}}{c_K(1-e_K)(\Gamma_{t_i}^0)^{e_K}+1}$, $F(K,\Gamma_{t_i}^0)=c_Ke_K(\Gamma_{t_i}^{0})^{e_K-1}$, and $G(K,\Gamma_{t_i}^0)=c_K(1-e_K)(\Gamma_{t_i}^{0})^{e_K}+1$ when $e_K>1$.

	By approximating $\tilde{\epsilon}_K(\Gamma_{t_i})$ with the proposed surrogate function in (\ref{eq: proposition 1 equation}), 
	the problem is transformed into a multiple-ratio FP problem. 
    \rv{Note that similar approximation method can be easily adopted for other transmission problems such as resource allocation by applying the proposed surrogate function in (\ref{eq:surrogate function}).}
	\rv{Moreover, this method can also be applied to multi-user multiple input multiple output (MU-MIMO) scenarios. 
    With the approximation method, the MU-MIMO beamforming problem can be reformulated as a fractional programming problem. 
    Then, we can either transform the MU-MIMO problem to MU-MISO problem using the method  in \cite{zhang2022deep} or directly solve the fractional programming problem with matrix variables to be optimized.}
    \subsubsection{Fractional Programming}
	In this part, we solve the multiple-ratio FP problem $\mathbf{P4}$. 
	Note that, $\mathbf{P4}$ still cannot be directly solved, 
	{because the QoS constraints are non-convex. Besides,  in terms of the beamforming vector, 
	both the objective and the constraints (\ref{eq: qos constraints in p3}) {are in fractional form}.} 
	To this end, the alternating optimization method is considered. 
	We first apply the Lagrangian dual {transformation} proposed in \cite{shen2018fractional} to (\ref{eq: qos constraints in p3}), and the problem can be equivalently written as 
    \vspace{-2mm}
	\begin{subequations}\label{eq: subproblem for beamforming optimization}
		\begin{align}
				\mathbf{P5}: \max_{\mathbf{V},\mathbf{y},\mathbf{z}}~~ &\sum_{i=1}^T \zeta_K\big(\mathbf{V},\Gamma_{t_i}^0\big),\\
							{\rm s.t.}~~&\frac{M_K}{L}
							R_{b_i,1}'\big(y_{i},\mathbf{V}\big)\notag\\&+ \frac{L-M_K}{L}R_{b_i,2}'\big(z_{i},\mathbf{V}\big)\geq \beta_{b_i}, \forall b_i \in \mathcal{B},\label{eq: yz added QOS constraints}
		\end{align}
		\end{subequations}
	where  $\mathbf{y} = [y_{1},...,y_{B}]^T\in \mathbb{R}^{B}$, $\mathbf{z}=[z_{1},...,z_{B}]^T\in \mathbb{R}^B$ are auxiliary variables for the SINR terms, 
	\vspace{-2mm}
    \begin{align}
		R_{b_i,1}'&=\log_2(1+y_{i})-y_{i}\notag\\&\qquad\qquad+{\frac{(1+y_{i})|\mathbf{h}_{b_i}^H\mathbf{v}_{b_i}|^2}{\sum_{j
					\in\left\{\mathcal{B},
					\mathcal{T}\right\} }\left|\mathbf{h}_{b_i}^H \mathbf{v}_{j}\right|^2+\frac{{\rm Tr}(\mathbf{V}\mathbf{V}^H)}{P_T}\sigma_{b_i}^2}},\\
		R_{b_i,2}'&=\log_2(1+z_{i})-z_{i}\notag\\&\qquad\qquad+\frac{(1+z_{i})\left|\mathbf{h}_{b_i}^H\mathbf{v}_{b_i}\right|^2}{\sum_{j
					\in\left\{\mathcal{B}\right\}}\left|\mathbf{h}_{b_i}^H \mathbf{v}_{j}\right|^2+\frac{{\rm Tr}(\mathbf{V}\mathbf{V}^H)}{P_T}\sigma_{b_i}^2}.
	\end{align}
	Note that, for maximizing $R_{b_i,1}'$ and $R_{b_i,2}'$ with a fixed $\mathbf{V}$, the optimal $\mathbf{y}$ and $\mathbf{z}$ equal to the corresponding SINR term of user $b_i$, as follows. 
	
	\vspace{-5mm}
	\begin{small}
		\begin{align}
			y_i &= \frac{|\mathbf{h}_{b_i}^H\mathbf{v}_{b_i}|^2}{\sum_{j\in\{\mathcal{B,T}\}/b_i}|\mathbf{h}_{b_i}^H\mathbf{v}_{j}|^2+\frac{{\rm Tr}(\mathbf{V}\mathbf{V}^H)}{P_T}\sigma_{b_i}^2},~i\in\{1,...,B\},\label{eq:update method of y}\\
			z_i &= \frac{|\mathbf{h}_{b_i}^H\mathbf{v}_{b_i}|^2}{\sum_{j\in\{\mathcal{B}\}/b_i}|\mathbf{h}_{b_i}^H\mathbf{v}_{j}|^2+\frac{{\rm Tr}(\mathbf{V}\mathbf{V}^H)}{P_T}\sigma_{b_i}^2},~~~~i\in\{1,...,B\}.\label{eq:update method of z}
			\end{align}
	\end{small}
	Moreover, {$R'_{b_i,1}=R_{b_i,1}$ and $R'_{b_i,2}=R_{b_i,2}$ hold for optimal $\{y_i\}_{i=1}^B$ and $\{z_i\}_{i=1}^B$, respectively.} 
	The same properties holds for $R_{b_i,2}^{'}$ and $\mathbf{z}$ as well. 
	Therefore, with the optimal $\mathbf{y}$ and $\mathbf{z}$, the problem $\mathbf{P5}$ can be reduced to 
	\vspace{-2mm}
	\begin{align}\label{p5: reduced QOS problem}
				\mathbf{P6}: \max_{\mathbf{V}} ~~ &\sum_{i=1}^T \zeta_K\big(\mathbf{V},\Gamma_{t_i}^0\big),\\
							{\rm s.t.}~~
							& (\ref{eq: yz added QOS constraints}).\notag
		\end{align}
		It can be found that in $\mathbf{P6}$, the sum-of-ratio form exists in both the objective and the QoS constraints. 
	{Therefore, we adopt the quadratic transformation proposed in \cite{shen2018fractional}, {which yields the following optimization problem.}} 
	
	\vspace{-5mm}
	\begin{small}
	\begin{subequations}\label{eq: subproblem for beamforming optimization}
            \begin{align}
                \mathbf{P7}: \max_{\mathbf{V},\mathbf{x},\mathbf{m},\mathbf{n}} ~~&\sum_{i=1}^TE(K,\gamma_{t_i}^0)\bigg[2x_{i}{\rm Re}\{\mathbf{h}_{t_i}^H\mathbf{v}_{t_i}\}\notag\\&-x_{i}^2\bigg(F(K,\gamma_{t_i}^0)|\mathbf{h}_{t_i}^H\mathbf{v}_{t_i}|^2+G(K,\gamma_{t_i}^0)\notag\\\cdot\big(&\sum_{j\in\{\mathcal{B,T}\}/t_i}|\mathbf{h}_{t_i}^H\mathbf{v}_j|^2+\frac{{\rm Tr}(\mathbf{V}\mathbf{V}^H)}{P_T}\sigma_{t_i}^2\big)\bigg)\bigg],\\
                {\rm s.t.}~~&\frac{M_K}{L}R_{b_i,1}''(\mathbf{V},y_{i},m_{i})\notag\\+&\frac{L-M_K}{L}R_{b_i,2}''(\mathbf{V},z_{i},n_{i})\geq \beta_{b_i}, \forall b_i \in \mathcal{B}, \label{eq: qos constraints after fractional programming}
            \end{align}
	\end{subequations}
\end{small}
	where $R_{b_i,1}''(\mathbf{V},y_{i},m_{i})$ and $R_{b_i,1}''(\mathbf{V},y_{i},m_{i})$ are given by 
	\vspace{-3mm}
    \begin{small}
		\begin{align}
			R_{b_i,1}''(\mathbf{V},y_{i},m_{i})&=\log\left(1+y_i\right)-y_i+ 2m_i\sqrt{1+y_i}{\rm Re}\{\mathbf{h}_{b_i}^H\mathbf{v}_{b_i}\}\notag\\&-m_i^2\big(\sum_{j\in \{\mathcal{B,T}\}}|\mathbf{h}_{b_i}^H\mathbf{v}_j|^2+\frac{{\rm Tr}(\mathbf{V}\mathbf{V}^H)}{P_T}\sigma_{b_i}^2\big),\\
			R_{b_i,2}''(\mathbf{V},z_{i},n_{i})&=\log\left(1+z_i\right)-z_i+ 2n_i\sqrt{1+z_i}{\rm Re}\{\mathbf{h}_{b_i}^H\mathbf{v}_{b_i}\}\notag\\&~~~-n_i^2\big(\sum_{j\in \{\mathcal{B}\}}|\mathbf{h}_{b_i}^H\mathbf{v}_j|^2+\frac{{\rm Tr}(\mathbf{V}\mathbf{V}^H)}{P_T}\sigma_{b_i}^2\big).
		\end{align}
	\end{small}
	In $\mathbf{P7}$, three auxiliary vectors $\mathbf{x}=[x_1,...,x_T]$, $\mathbf{m}=[m_1,...,m_B]$, and $\mathbf{n}=[n_1,...,n_B]$ are introduced to transform the original problem to a quadratic programming problem.  
	{More specifically, with a given 
    $\mathbf{V}$, 
	the optimal $\mathbf{x}$, $\mathbf{m}$, and $\mathbf{n}$ for $\mathbf{P7}$ are given as follows.} 
	
	\vspace{-5mm}
	\begin{small}
		\begin{align}
			x_{i}&={\rm Re}\{\mathbf{h}_{t_i}^H\mathbf{v}_{t_i}\}\bigg[F(K,\Gamma_{t_i}^0)|\mathbf{h}_{t_i}^H\mathbf{v}_{t_i}|^2+G(K,\Gamma_{t_i}^0)\notag\\\cdot&{\big(\sum_{j\in\{\mathcal{B,T}\}/t_i}|\mathbf{h}_{t_i}^H\mathbf{v}_j|^2+\frac{{\rm Tr}(\mathbf{V}\mathbf{V}^H)}{P_T}\sigma_{t_i}^2\big)}\bigg]^{-1},\forall i=\{1,...,T\},  \label{eq: update method of x} \\
			m_i&=\frac{\sqrt{1+y_i}{\rm Re}\{\mathbf{h}_{b_i}^H\mathbf{v}_{b_i}\}}{\sum_{j\in \{\mathcal{B,T}\}}|\mathbf{h}_{b_i}^H\mathbf{v}_j|^2+\frac{{\rm Tr}(\mathbf{V}\mathbf{V}^H)}{P_T}\sigma_{b_i}^2},\forall i=\{1,...,B\}, \label{eq: update method of m} \\
			n_i&=\frac{\sqrt{1+z_i}{\rm Re}\{\mathbf{h}_{b_i}^H\mathbf{v}_{b_i}\}}{\sum_{j\in \{\mathcal{B}\}}|\mathbf{h}_{b_i}^H\mathbf{v}_j|^2+\frac{{\rm Tr}(\mathbf{V}\mathbf{V}^H)}{P_T}\sigma_{b_i}^2},\forall i=\{1,...,B\}. \label{eq: update method of n} 	
		\end{align}
	\end{small}Then, in terms of fixed $\mathbf{x}$, $\mathbf{m}$, and $\mathbf{n}$, the problem $\mathbf{P6}$ can be further reduced to 
	\vspace{-2mm}
	\begin{small}
    \begin{subequations}\label{prob: final qcqp problem}
		\begin{align}
		\mathbf{P8}: \max_{\mathbf{V}} &\sum_{i=1}^TE(K,\gamma_{t_i}^0)\bigg[2x_{t_i}{\rm Re}\{\mathbf{h}_{t_i}^H\mathbf{v}_{t_i}\}\notag\\&-x_{t_i}^2\bigg(F(K,\Gamma_{t_i}^0)|\mathbf{h}_{t_i}^H\mathbf{v}_{t_i}|^2+G(K,\Gamma_{t_i}^0)\notag \\ &\big(\sum_{j\in\{\mathcal{B,T}\}/t_i}|\mathbf{h}_{t_i}^H\mathbf{v}_j|^2+\frac{{\rm Tr}(\mathbf{V}\mathbf{V}^H)}{P_T}\sigma_{t_i}^2\big)\bigg)\bigg],\\
		{\rm s.t.}~~&(\ref{eq: qos constraints after fractional programming}). \notag
	\end{align}
	\end{subequations}
\end{small}It can be observed that $\mathbf{P8}$ is actually an inhomogeneous and separable QCQP problem,  
	which can be solved by convex optimization toolboxes like CVX in MATLAB. 
	However, the complexity of CVX is still unbearable since $\mathbf{P8}$ needs to be solved in each iteration. 
	Therefore, 
    {we derive a semi-closed form solution for $\mathbf{P8}$ and propose a computationally efficient fixed point algorithm to search for the Lagrangian multipliers. 
    Formally,} 
    the Lagrangian of $\mathbf{P8}$ is given by 

	\vspace{-4mm}
    \begin{small}
        \begin{align}\label{eq: Lagrangian function of problem p8}
            \mathcal{L}(\mathbf{V},\boldsymbol{\lambda})=&\sum_{i=1}^TE(K,\gamma_{t_i}^0)\bigg[2x_{t_i}{\rm Re}\{\mathbf{h}_{t_i}^H\mathbf{v}_{t_i}\}\notag\\&-x_{t_i}^2\bigg(F(K,\Gamma_{t_i}^0)|\mathbf{h}_{t_i}^H\mathbf{v}_{t_i}|^2+G(K,\Gamma_{t_i}^0)\notag\\ &\cdot \big(\sum_{j\in\{\mathcal{B,T}\}/t_i}|\mathbf{h}_{t_i}^H\mathbf{v}_j|^2+\frac{{\rm Tr}(\mathbf{V}\mathbf{V}^H)}{P_T}\sigma_{t_i}^2\big)\bigg)\bigg]\notag\\
            &+\sum_{i=1}^B \lambda_i\bigg[\frac{M_K}{L}R_{b_i,1}''(\mathbf{V},y_{i},m_{i})\notag \\&+\frac{L-M_K}{L}R_{b_i,2}''(\mathbf{V},z_{i},n_{i})- \beta_{b_i}\bigg],
        \end{align}
    \end{small}where $\boldsymbol{\lambda}=[\lambda_1,...,\lambda_B]$ is the vector composed of multiple non-negative Lagrange multipliers. 

	By taking the first-order derivative of $\mathcal{L}$ over the precoding vectors (i.e., $\{\mathbf{v}_{t_i}\}_{i=1}^T$, $\{\mathbf{v}_{b_i}\}_{i=1}^B$) and setting it to zero, 
	we have the following proposition. 
	\vspace{-0.09in}
	\begin{proposition}\label{prop: the optimal solution of p8}
		\emph{For the MU-MISO system with channel $\{\{\mathbf{h}_{b_i}\}_{i=1}^B,\{\mathbf{h}_{t_i}\}_{i=1}^T\}$, the optimal solution of the problem $\mathbf{P8}$ is given by} 
        \vspace{-2mm}
		\begin{align}
			\mathbf{v}_{b_i}&=\underbrace{\lambda_i\frac{M_Km_i\sqrt{1+y_i}+(L-M_K)n_i\sqrt{1+z_i}}{L}}_{\rho_{b_i}}\mathbf{A}^{-1}\mathbf{h}_{b_i}\label{eq: update of bituser precoding with lambda},\\ 
			&{\rm where}~\mathbf{A}=\frac{\sum_{j=1}^B\mu_j\sigma_{b_j}^2+\sum_{j=1}^T\nu_j\sigma_{t_j}^2}{P_T}\mathbf{I}+\sum_{j=1}^B \mu_j\mathbf{h}_{b_j}\mathbf{h}_{b_j}^H \notag\\&\qquad\qquad\qquad\qquad\qquad\qquad\qquad+\sum_{j=1}^T\nu_j\mathbf{h}_{t_j}\mathbf{h}_{t_j}^H,\label{eq: calculate of A}
			\\\mathbf{v}_{t_i}&=\underbrace{x_iE(K,\Gamma_{t_i}^0)}_{\varsigma_{t_i}}\mathbf{B}_{t_i}^{-1}\mathbf{h}_{t_i}\label{eq: update of semuser precoding with lambda}, \vspace{-10mm}\\
			&{\rm where}~\mathbf{B}_{t_i}=\frac{\sum_{j=1}^B\mu_j\sigma_{b_j}^2+\sum_{j=1}^T\nu_j\sigma_{t_j}^2}{P_T}\mathbf{I}\notag\\&+x_i^2E(K,\Gamma_{t_i}^0)F(K,\Gamma_{t_i}^0)\mathbf{h}_{t_i}\mathbf{h}_{t_i}^H+\sum_{j=1}^B\frac{\lambda_jM_K}{L}m_j^2\mathbf{h}_{b_j}\mathbf{h}_{b_j}^H\notag\\&\qquad\qquad\qquad\qquad\qquad\qquad+\sum_{j=1,j\neq t_i}^T\nu_j\mathbf{h}_{t_j}\mathbf{h}_{t_j}^H, \label{eq: calculate of B}
		\end{align}
		\emph{where $\mu_{j}=\frac{m_j^2\lambda_jM_K+n_j^2\lambda_j(L-M_K)}{L}$, $\nu_j = x_j^2E(K,\Gamma_{t_j}^0)G(K,\Gamma_{t_j}^0)$. }	
	\end{proposition}
	\vspace{-4mm}
	\begin{remark}\label{remark: discussion about the proposition 1}
		\emph{(Optimal Beamforming Structure) 
		As shown in Proposition 1, it can be observed that the precoding vectors of both bit-users and sem-users are linear transformations of their corresponding channel vectors, where the weight coefficients is divided into linear power allocation coefficients ({i.e., $\rho_{b_i}$ and $\varsigma_{t_i}$ in (\ref{eq: update of bituser precoding with lambda}) and (\ref{eq: update of semuser precoding with lambda})}) and priority coefficients for interference suppression ({i.e., $\mu_j$, $\nu_j$, $\frac{\lambda_jM_K}{L}m_j^2$ in $\mathbf{A}$ and $\{\mathbf{B}_{t_i}\}_{i=1}^T$}). Comparing (\ref{eq: update of bituser precoding with lambda}) and (\ref{eq: update of semuser precoding with lambda}), we can find that $\mathbf{v}_ {t_i} $and $\mathbf{v}_ {b_i} $ have different weight coefficients, indicating that sem-users have different resource allocation strategies compared to traditional digital communication due to their different objective functions. Furthermore, according to Proposition 1, the optimal precoding vector of $\mathbf{P8}$ is only determined by $ \{\lambda_i\}_ {i=1} ^ B$, so we can turn to find the optimal $ \{\lambda_i\}_ {i=1} ^ B$ for solving $\mathbf{P8}$, thereby reducing computational complexity. 
		}
	\end{remark}
	\begin{remark}
		\emph{(Computation Complexity Analysis) 
		For the calculation of $\{\mathbf{v}_{b_i}\}_{i=1}^B$, it can be observed that the inverse matrix is shared among bit-users, thus only needs to be calculated once. The complexity for calculating $\{\mathbf{v}_{b_i}\}_{i=1}^B$ is given by $\mathcal{O}(BN_t^{2}+TN_t^{2}+N_t^{3})$. 
		For the calculation of $\{\mathbf{v}_{t_i}\}_{i=1}^T$, according to the Sherman–Morrison formula, we have $\mathbf{B}_{t_i}= (\mathbf{C}-\omega_{t_i}\mathbf{h_{t_i}\mathbf{h}_{t_i}^H})^{-1}=\mathbf{C}^{-1}+\omega_{t_i}\frac{\mathbf{C}^{-1}\mathbf{h}_{t_i}\mathbf{h}_{t_i}^H\mathbf{C}^{-1}}{1-\omega_{t_i}\mathbf{h}_{t_i}^H\mathbf{C}^{-1}\mathbf{h}_{t_i}}$, where $\mathbf{C}$ is defined in (\ref{eq: the common inverse matrix of semusers}), and $\omega_{t_i}=\nu_i-x_i^2E(K,\Gamma_{t_i}^0)F(K,\Gamma_{t_i}^0)$. 
		Therefore, the complexity for calculating $\{\mathbf{v}_{t_i}\}_{i=1}^T$ is given by $\mathcal{O}(BN_t^2+TN_t^2+N_t^3)$. 
		In a nutshell, the overall complexity for calculating the beamforming vectors by (\ref{eq: update of bituser precoding with lambda}) and (\ref{eq: update of semuser precoding with lambda}) is $\mathcal{O}(BN_t^2+TN_t^2+N_t^3)$. 
		}
	\end{remark}
	\begin{figure}[t]
        \label{alg:LSB}
		\vspace{-2mm}
        \begin{algorithm}[H]
          \caption{The proposed method for solving $\mathbf{P8}$}\label{algo:method for solving p8}
          \begin{algorithmic}[1]
              \renewcommand{\algorithmicrequire}{ \textbf{Initialize}}
              \REQUIRE $\boldsymbol{\lambda}=[\lambda_{1},...,\lambda_{B}]=0.01\times \mathbf{1}, \boldsymbol{\lambda}'=\mathbf{0}$. 
			\REPEAT
			\STATE Update $\boldsymbol{\lambda}'$ by $\boldsymbol{\lambda}' \leftarrow \boldsymbol{\lambda}$. 
			\STATE Update precoding matrix $\mathbf{V}$ by (\ref{eq: update of bituser precoding with lambda}), (\ref{eq: update of semuser precoding with lambda}), and $\{\lambda_i\}_{i=1}^B$.  
			\STATE Update  $\{\lambda_i\}_{i=1}^B$ by (\ref{eq: update of lambda}). 
			\UNTIL {$\sum_{i=1}^B |\lambda_i -\lambda_i'|\leq \xi$} 
			\renewcommand{\algorithmicrequire}{ \textbf{Output}}
			\REQUIRE the precoding matrix $\mathbf{V}$. 
          \end{algorithmic}
        \end{algorithm}
		\vspace{-6mm}
      \end{figure}
	  \begin{figure}[t]
		\vspace{-5mm}
		  \label{alg:LSB}
		  \begin{algorithm}[H]
			\caption{The proposed MM-FP algorithm for solving $\mathbf{P2}$}\label{algo:sca fp algorithm}
			\begin{algorithmic}[1]
				\renewcommand{\algorithmicrequire}{ \textbf{Initialize}}
				\REQUIRE the precoding matrix $\mathbf{V}$. 
				\REPEAT
				\STATE Update the SINR related terms, i.e., $\Gamma_{t_i,0}\leftarrow \Gamma_{t_i}$, $\mathbf{y}$ with (\ref{eq:update method of y}), $\mathbf{z}$ with (\ref{eq:update method of z}). 
				\STATE Update the ratio related terms, i.e.,  $\mathbf{x}$ with (\ref{eq: update method of x}), $\mathbf{m}$ with (\ref{eq: update method of m}), $\mathbf{n}$ with (\ref{eq: update method of n}). 
				\STATE Update the beamforming matrixes $\mathbf{V}$ by solving $\mathbf{P8}$ with Algorithm \ref{algo:method for solving p8}. 
				\UNTIL the objective value of $\mathbf{P2}$ converges or the iteration number reaches the maximum. 
				\STATE Normalize the precoding matrix, i.e., $\mathbf{V}\leftarrow \sqrt{\frac{P_T}{{\rm Tr}(\mathbf{V}\mathbf{V}^H)}}\mathbf{V}$. 
			\end{algorithmic}
		  \end{algorithm}
		  \vspace{-10mm}
		\end{figure}
	  \begin{figure*}[!]
		\begin{small}
		\begin{align}
			\mathbf{C}&=\frac{\sum_{j=1}^B\mu_j\sigma_{b_j}^2+\sum_{j=1}^T\nu_j\sigma_{t_j^2}}{P_T}\mathbf{I}+\sum_{j=1}^B\frac{\lambda_jM_K}{L}m_j^2\mathbf{h}_{b_j}\mathbf{h}_{b_j}^H+\sum_{j=1}^T\nu_j\mathbf{h}_{t_j}\mathbf{h}_{t_j}^H,\label{eq: the common inverse matrix of semusers}\\
			\lambda_i&=\frac{{\beta_{b_i}-\frac{M_K}{L}(\log(1+y_i)-y_i-m_i^2\frac{{\rm Tr}(\mathbf{V}\mathbf{V}^H)}{P_T}\sigma_{b_i}^2)-\frac{L-M_K}{L}(\log(1+z_i)-z_i-n_i^2\frac{{\rm Tr}(\mathbf{V}\mathbf{V}^H)}{P_T}\sigma_{b_i}^2)}}{{2(\frac{M_K}{L}m_i\sqrt{1+y_i}+\frac{L-M_K}{L}n_i\sqrt{1+z_i})^2}{\rm Re}\{\mathbf{h}_{b_i}^H\mathbf{A}^{-1}\mathbf{h}_{b_i}\}}\notag\\&\qquad\qquad\qquad\qquad\qquad\qquad\qquad+\frac{\frac{M_K}{L}m_i^2\big(\sum_{j\in \{\mathcal{B,T}\}}|\mathbf{h}_{b_i}^H\mathbf{v}_j|^2\big) + \frac{L-M_K}{L}n_i^2\big(\sum_{j\in \{\mathcal{B}\}}|\mathbf{h}_{b_i}^H\mathbf{v}_j|^2\big)}{{2(\frac{M_K}{L}m_i\sqrt{1+y_i}+\frac{L-M_K}{L}n_i\sqrt{1+z_i})^2}{\rm Re}\{\mathbf{h}_{b_i}^H\mathbf{A}^{-1}\mathbf{h}_{b_i}\}}.\label{eq: update of lambda}
		\end{align}
		\end{small}
		\vspace{-2mm}
			\hrule
		\vspace{-7mm}
	\end{figure*}
	As discussed in Remark \ref{remark: discussion about the proposition 1}, to obtain the optimal solution of $\mathbf{P8}$, only the Lagrange multipliers $\{\lambda_i^*\}_{i=1}^B$ needs to be determined, where $\lambda_i^*$ denotes the optimal dual variable.  
{	In addition, it can be found that 
	the optimal solution of $\mathbf{P8}$ should satisfy the QoS constraints with equality. 
}	
	{As a result, we can obtain  $\{\lambda_i^*\}_{i=1}^B$ by the fixed-point algorithm.}
	The update {rule} is presented in (\ref{eq: update of lambda}). 
	The solution process for $\mathbf{P8}$ is concluded in Algorithm \ref{algo:method for solving p8}. 	

	So far, the problem $\mathbf{P2}$ has been solved in an alternating manner, {which is summarized} in Algorithm \ref{algo:sca fp algorithm} and termed as majorization-minimization fractional programming (MM-FP). 
	There are two groups of introduced variables, namely SINR related terms ($\{\gamma_{t_i,0}\}_{i=1}^T$, $\mathbf{y}$, $\mathbf{z}$) and the {fraction} related terms ($\mathbf{x}$, $\mathbf{m}$, $\mathbf{n}$). 
	As shown in (\ref{eq: update method of x})-(\ref{eq: update method of n}), 
    the two groups are updated in sequence, followed by the update of $\mathbf{V}$. 
\rv{Algorithm \ref{algo:sca fp algorithm} requires multiple iterations, each of which is divided into three steps. The first step includes the update of SINR related terms, with a complexity of $\mathcal{O}(BN_t+TN_t)$; the second step includes the updates of the ratio related terms, with a complexity of $\mathcal{O}(BN_t+TN_t)$; the third step includes the updates of precoding matrix, with a complexity of $\mathcal{O}(L_2(BN_t^2+TN_t^2+N_t^3))$, where $L_2$ denotes the the number of iteration rounds for the fixed point method. 
Therefore, the complexity of Algorithm \ref{algo:sca fp algorithm} is 
\begin{align}\label{eq:computation complexity of MM-FP algorithm}
\mathcal{O}(L_1L_2BN_t^2+L_1L_2TN_t^2+L_1N_t^3), 
\end{align}
where $L_1$ denotes the number of iterations. 
}

 To address concerns about the high computational complexity introduced by multiple iterations, we also propose a low-complexity beamforming method to enhance practicality. 
 Generally, beamforming involves determining the precoding direction and allocating power. As shown in (34) and (36),  the precoding direction is determined by the channel vectors, $\{m_i\}_{I=1}^B$, $\{n_i\}_{I=1}^B$, $\{x_i\}_{I=1}^B$, and $\{\lambda_i\}_{I=1}^B$. Given this, we consider setting a uniform $\lambda_i$ for all bit-user $b_i$ for avoiding iteration, and approximating the value of $\{m_i\}_{I=1}^B$, $\{n_i\}_{I=1}^B$, $\{x_i\}_{I=1}^B$ by (29), (30), (31), where the precoding vectors are replaced by the corresponding channel vector. Consequently, the beamforming direction, denoted by $\{\widetilde{v}_{b_i}\}_{I=1}^B,\{\widetilde{v}_{t_i}\}_{I=1}^T$, is established. Then we allocate the power by solving the following problem. 

	\begin{small}
	\begin{subequations}\label{eq:power allocation subproblem}
			\begin{align}
				&\mathbf{P9}:\notag \\ 
				&\max_{\mathbf{p}_T,\mathbf{p}_B}\sum_{i=1}^T a_K+\frac{d_K}{c_K+(\frac{p_{t_i}|\mathbf{h}_{t_i}^H\widetilde{\mathbf{v}}_{t_i}|^2}{(\sum_{j\in \{\mathcal{B,T}\}/t_i }p_{j}|\mathbf{h}_{t_i}^H\widetilde{\mathbf{v}}_j|^2+\sigma_{t_i}^2)})^{-e_K}},\label{eq:opjective function in p10}\\
				&{\rm s.t.} \frac{M_K}{L}\log_2\left(1+\frac{p_{b_i}\left|\mathbf{h}_{b_i}^H\widetilde{\mathbf{v}}_{b_i}\right|^2}{\sum_{j
				\in\left\{\mathcal{B},
				\mathcal{T}\right\}/b_i }p_{j}\left|\mathbf{h}_{b_i}^H\widetilde{\mathbf{v}}_{j}\right|^2+\sigma_{b_i}^2}\right)\notag + \frac{L-M_K}{L}\\ &~~~\cdot\log_2\left(1+\frac{p_{b_i}\left|\mathbf{h}_{b_i}^H\widetilde{\mathbf{v}}_{b_i}\right|^2}{\sum_{j
				\in\left\{\mathcal{B}\right\}/b_i }p_{j}\left|\mathbf{h}_{b_i}^H \widetilde{\mathbf{v}}_{j}\right|^2+\sigma_{b_i}^2}\right) \geq \beta_{b_i}, \forall b_i,\label{eq: qos constraints in p10} \\
				&~~~~\sum_{i=1}^Bp_{b_i} + \sum_{j=1}^Tp_{t_j} \leq P_T. \label{eq:power constraints in p10}
				\end{align}
	\end{subequations}
\end{small}where $\mathbf{p}_T=[p_{t_1},...,p_{t_T}]$ and $\mathbf{p}_B=[p_{b_1},...,p_{b_B}]$ denote the power allocation vectors of sem-users and bit-users, respectively. 

	The low-complexity majorization-minimization fractional programming (LP-MM-FP) algorithm is concluded in Algorithm \ref{algo:low complexity algorithm for solving the original problem}. 
\rv{The first step includes the calculation of ratio and SINR related terms, with a complexity of $\mathcal{O}(BN_t+TN_t)$; the second step includes determining beamforming direction, with a complexity of $\mathcal{O}(BN_t^2+TN_t^2+N_t^3)$; the third step is conducted for power allocation, with a complexity of $\mathcal{O}((B+T)^{3.5})$. The overall complexity of Algorithm \ref{algo:low complexity algorithm for solving the original problem} is given by
\begin{align}\label{eq: computation complexity of LP-MM-FP}
\mathcal{O}(BN_t^2+TN_t^2+N_t^3+(B+T)^{3.5}).
\end{align}
Comparing (\ref{eq: computation complexity of LP-MM-FP}) with (\ref{eq:computation complexity of MM-FP algorithm}), the number of users  $B+T$  is generally much smaller than the number of antennas. 
This suggests that Algorithm \ref{algo:low complexity algorithm for solving the original problem} has a lower computational complexity compared to Algorithm \ref{algo:sca fp algorithm}.
}
	\begin{figure}[t]
		\vspace{-2mm}
        \label{alg:LSB}
        \begin{algorithm}[H]
          \caption{The proposed LP-MM-FP algorithm for solving $\mathbf{P2}$}\label{algo:low complexity algorithm for solving the original problem} 
          \begin{algorithmic}[1]			
			\STATE $\lambda_i\leftarrow \lambda_0$, $i=1,...,B$.
			\STATE Calculate $\mathbf{m}$, $\mathbf{n}$, $\mathbf{x}$ ,$\{\Gamma_{t_i}\}_{i=1}^T$ by channel vectors and MRT precoding vectors (i.e., $\mathbf{v}_{t_i}=\frac{\mathbf{h}_{t_i}}{\|\mathbf{h}_{t_i}\|},\mathbf{v}_{b_i}=\frac{\mathbf{h}_{b_i}}{\|\mathbf{h}_{t_i}\|}$). 
			\STATE Calculate the beamforming directions with $\{\lambda_i\}_{i=1}^B$ and $\{\Gamma_{t_i}\}_{i=1}^T$,  i.e., $\widetilde{\mathbf{v}}_{t_i}=\frac{\mathbf{B}_{t_i}^{-1}\mathbf{h}_{t_i}}{\|\mathbf{B}_{t_i}^{-1}\mathbf{h}_{t_i}\|}$, $\widetilde{\mathbf{v}}_{t_i}=\frac{\mathbf{A}^{-1}\mathbf{b}_{t_i}}{\|\mathbf{A}^{-1}\mathbf{h}_{b_i}\|}$, where $\mathbf{A}$, $\mathbf{B}_{t_i}$ are given by (\ref{eq: calculate of A}) and (\ref{eq: calculate of B}), respectively. 
			\STATE Calculate power allocation vectors $\mathbf{p}_T$ and $\mathbf{p}_B$ by solving problem $\mathbf{P9}$ with $\{\widetilde{\mathbf{v}}_{t_i}\}_{i=1}^T$ and $\{\widetilde{\mathbf{v}}_{b_i}\}_{i=1}^B$. 
			\STATE Calculate the final precoding vectors by $\mathbf{v}_{t_i}=\sqrt{p_{t_i}}\widetilde{\mathbf{v}}_{t_i}$, $\mathbf{v}_{b_i}=\sqrt{p_{b_i}}\widetilde{\mathbf{v}}_{b_i}$. 
          \end{algorithmic}
        \end{algorithm}
		\vspace{-9mm}
      \end{figure}
      \vspace{-1mm}
      \begin{figure}[b]
          \label{alg:LSB}
		  \vspace{-8mm}
          \begin{algorithm}[H]
            \caption{The proposed beamforming algorithm for solving $\mathbf{P1}$}\label{algo:sca-fp algorithm for solving the original problem} 
            \begin{algorithmic}[1]
              \renewcommand{\algorithmicrequire}{ \textbf{Input}}
              \REQUIRE The sample set $\mathcal{K}=\{K|K_{\rm min}\leq K\leq K_{\rm max}, K\in \mathbb{Z}\}$
              \FOR {$K=K_{\min}, ..., K_{\rm max}$ \textbf{in parallel}} 
              \STATE Obtain $\mathbf{V}_{K}$ through Algorihtm \ref{algo:sca fp algorithm} with $K$ or Algorithm \ref{algo:low complexity algorithm for solving the original problem} with $K$. 
              \STATE Calculate the objective value of problem denoted by $\mathbf{P1}$ $\tau(\mathbf{V}_K, K)$ with $\mathbf{V}_{K}$ and $K$.
              \ENDFOR
              \STATE $K_{\rm opt}=\arg\max_{K\in \mathcal{K}}{\tau(\mathbf{V}_K,K)}$
              \renewcommand{\algorithmicrequire}{ \textbf{Output}}
              \REQUIRE The beamforming matrix $\mathbf{V}_{K_{\rm opt}}$, $K_{\rm opt}$. 
            \end{algorithmic}
          \end{algorithm}
          \vspace{-10mm}
        \end{figure}
      \subsection{Overall Algorithm}
	  \begin{figure*}[h]
		\centering
		\subfigure[Low SNR case (${\rm SNR}=0$ dB)]{
			\label{Fig:system comparison: low snr case} 
			\includegraphics[width=0.45\linewidth]{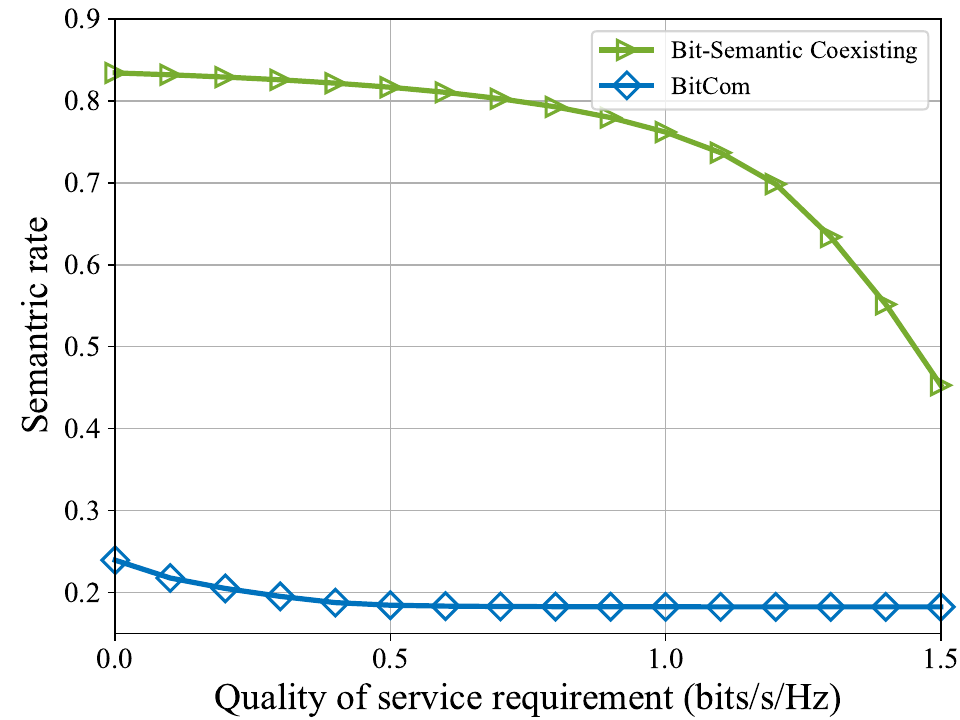}}
		\vspace{-3mm}
			\subfigure[High SNR case (${\rm SNR}=5$ dB)]{
			\label{Fig:system comparison: high snr case} 
			\includegraphics[width=0.45\linewidth]{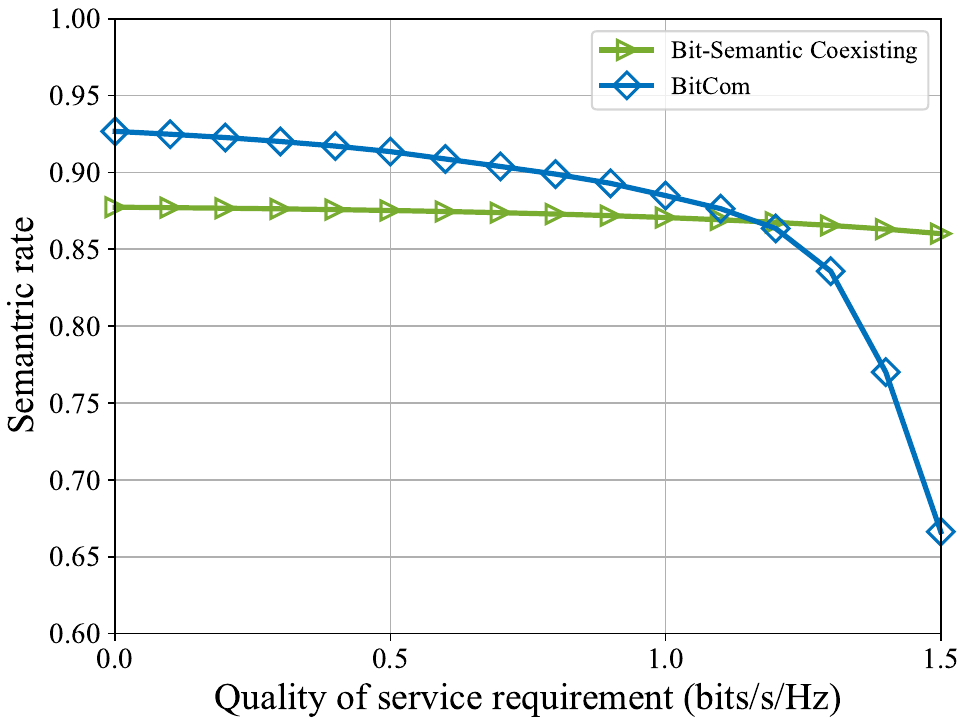}}
		\vspace{4mm}
			\caption{Performance comparison of transmission systems}\label{Fig:transmission system comparison}
		\vspace{-4mm}
	\end{figure*}
	In this subsection, we shall present the proposed method for solving the problem $\mathbf{P1}$. Firstly, the subproblem that optimizes the beamforming vector with a fixed $K$ has been tackled in a computation-efficient manner. 
	Subsequently, recognizing that the feasible set for $K$ (i.e., the downsampling depth) is typically constrained within a narrow integer range, we employ the exhaustive algorithm to identify the optimal $K$. 

    The overall algorithm is presented in Algorithm \ref{algo:sca-fp algorithm for solving the original problem}. 
It can be found that executing Algorithm \ref{algo:sca-fp algorithm for solving the original problem} requires at most $K_{\rm max}-K_{\rm min}$ times the complexity of Algorithm \ref{algo:sca fp algorithm}, where $K_{\rm max}$ and $K_{\rm min}$  represent the maximum and minimum feasible values of $K$, respectively. 
For conducting Algorithm \ref{algo:sca fp algorithm}, it requires multiple iterations, each of which is divided into three steps. The first step includes the update of SINR related terms, with a complexity of $\mathcal{O}(BN_t+TN_t)$; the second step includes the updates of the ratio related terms, with a complexity of $\mathcal{O}(BN_t+TN_t)$; the third step includes the updates of precoding matrix, with a complexity of $\mathcal{O}(L_2(BN_t^2+TN_t^2+N_t^3))$, where $L_2$ denotes the the number of iteration rounds for the fixed point method. Therefore, the complexity for conducting Algorithm \ref{algo:sca fp algorithm} is $\mathcal{O}(L_1L_2BN_t^2+L_1L_2TN_t^2+L_1N_t^3)$, where $L_1$ denotes the iteration number of Algorithm \ref{algo:sca fp algorithm}. 
In conclusion, the complexity for conducting Algorithm \ref{algo:sca-fp algorithm for solving the original problem} is given by $\mathcal{O}((K_{\rm max}-K_{\rm min})L_1L_2BN_t^2+(K_{\rm max}-K_{\rm min})L_1L_2TN_t^2+(K_{\rm max}-K_{\rm min})L_1N_t^3)$. 
\vspace{4mm}
	\section{Numerical Results}\label{sec: experiments}

	\subsection{Simulation Setup}
	\vspace{-1mm}
	\textbf{System setup.} We consider the clustered Saleh-Valenzuela channel model \cite{gao2023deep}, {in which} the channel from the BS to a specific user $i$ is given as follows. 
    \vspace{-2mm}
	\begin{align}
		\mathbf{h}_i = \frac{1}{\sqrt{L_p}}\sum_{l=1}^{L_p}\delta_{i,l} \mathbf{a}_T(\theta_{i,l}), 
	\end{align}
	{where $L_p$ is the number of paths and $\delta_{i,l}\sim\mathcal{CN}(0,1)$ is the channel attenuation of the $l$-th path.} Without loss of generality, we set $L_p$ to $10$.  
	$\theta_{i,l}$ denotes the azimuth angle of departure (AoD) at the transmitter, and we assume $\theta_{i,l}$ follows a uniform distribution from $0$ to $2\pi$. 
	The response vector of Uniform Linear Array (ULA) at the BS side can be expressed as
	\vspace{-1mm}
    \begin{align}
	\mathbf{a}_{T}(\theta_{i,l})=[1,e^{-j\pi \sin(\theta_{i,l})},...,e^{-j\pi(N_t-1)\sin(\theta_{i,l})}].
	\end{align}

	For the MISO system setting, unless specified, the following system parameters will be used as the default setting in the experiments: {$N_t=16$, $B=5$, $T=3$, $K=3$, ${\rm SNR}=0$ dB, $\beta_i = 1, \forall i$}.\footnote{{It is worth noting that the proposed method should be also adaptable to other configurations.}} 
	Besides, as mentioned in Section \ref{subsec:semantic rate approximation}, the ImageNet dataset and  SSIM are used as the training dataset and performance metric, respectively. 
	\rv{The image size $I$ is set to $128$.
	The length of the frame $L$ is set to $32,768$, and the number of filters $C$ is set to $128$.
	The feasible set of downsampling depth is given by $K\in \{K|2\leq K\leq 6, K\in \mathbb{Z}\}$. 
	Using (\ref{eq: calculation of M}), the corresponding feasible set of $M_K$ is given by $\{128,512,2048,8192,32768\}$.}
	We also conduct performance evaluation on the Kodak image dataset\footnote{\url{https://r0k.us/graphics/kodak/}}, 
    which comprises of 24 high-quality images. 

	\textbf{Benchmark schemes.} 
	We compare the proposed beamforming algorithm, MM-FP and LP-MM-FP, with three commonly-adopted beamforming schemes, including the zero focing (ZF) algorithm, maximum ratio transmission (MRT) algorithm, and {weighted minimum mean-square error} (WMMSE) algorithm. 
	Note that the aforementioned algorithms cannot be directly used for solving $\mathbf{P1}$, as the QoS constraints may not be satisfied. 
	Given this, we first obtain the beamforming direction through these algorithms, i.e., $\widetilde{\mathbf{v}}_i={\mathbf{v}_i}/{\|\mathbf{v}_i\|_2}, \forall i\in \mathcal{B}\cup \mathcal{T}$. 
	Then we reallocate the power by solving the problem $\mathbf{P9}$, and the final beamforming vector is given by $\mathbf{v}_{i}=\sqrt{p_i}\widetilde{\mathbf{v}}_i, \forall i\in \mathcal{B}\cup \mathcal{T}$. 
	{The resulting benchmark schemes are named ZF-PC, MRT-PC, and WMMSE-PC respectively.} 
    \vspace{-4mm}
	\begin{figure}[b]
		\centering
		\includegraphics[width=0.9\linewidth]{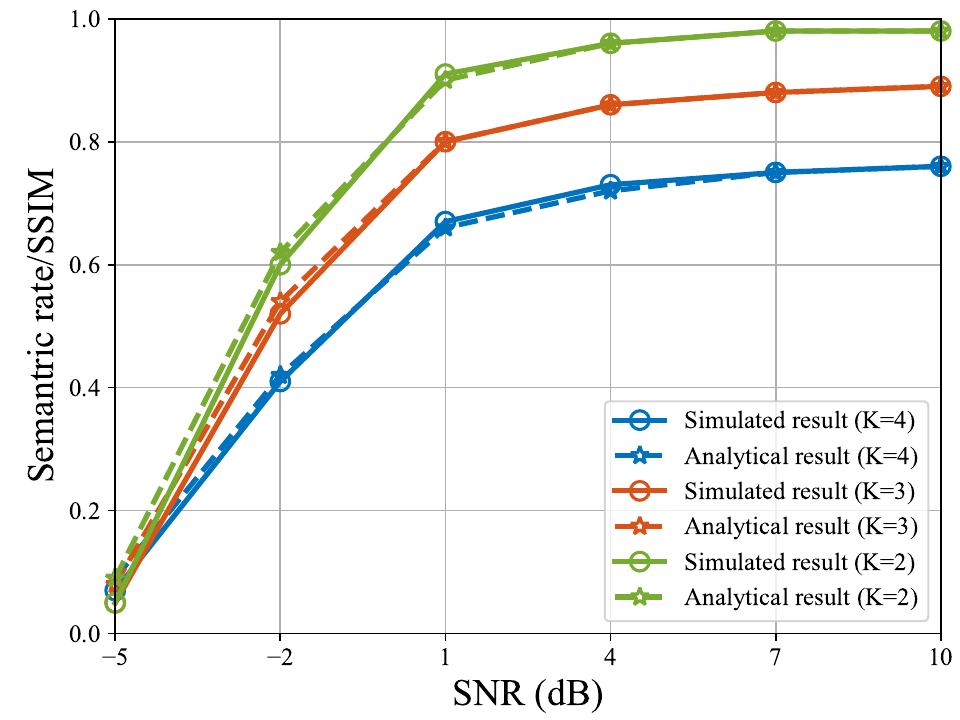}
		\vspace{1mm}
		\caption{{Validation of Semantic Rate Approximation}}\label{fig: theoretical validation}
		\vspace{-8mm} 
		\label{Fig:AD test acc}
	\end{figure}
	\begin{figure}[h]
		\centering
		\subfigure[${\rm SNR=0}$ dB]{
			\label{Fig:beamforming different qos} 
			\includegraphics[width=0.9\linewidth]{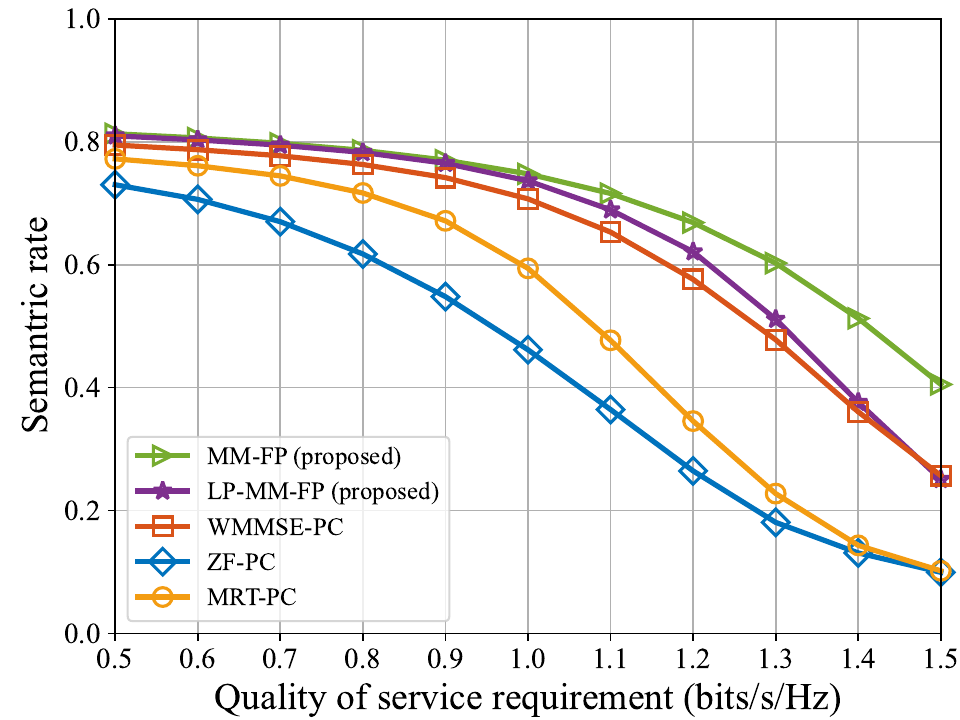}}
		\subfigure[$\beta_{i}=1, \forall i=1,...,B$]{
			\label{Fig:beamforming different snr} 
			\includegraphics[width=0.9\linewidth]{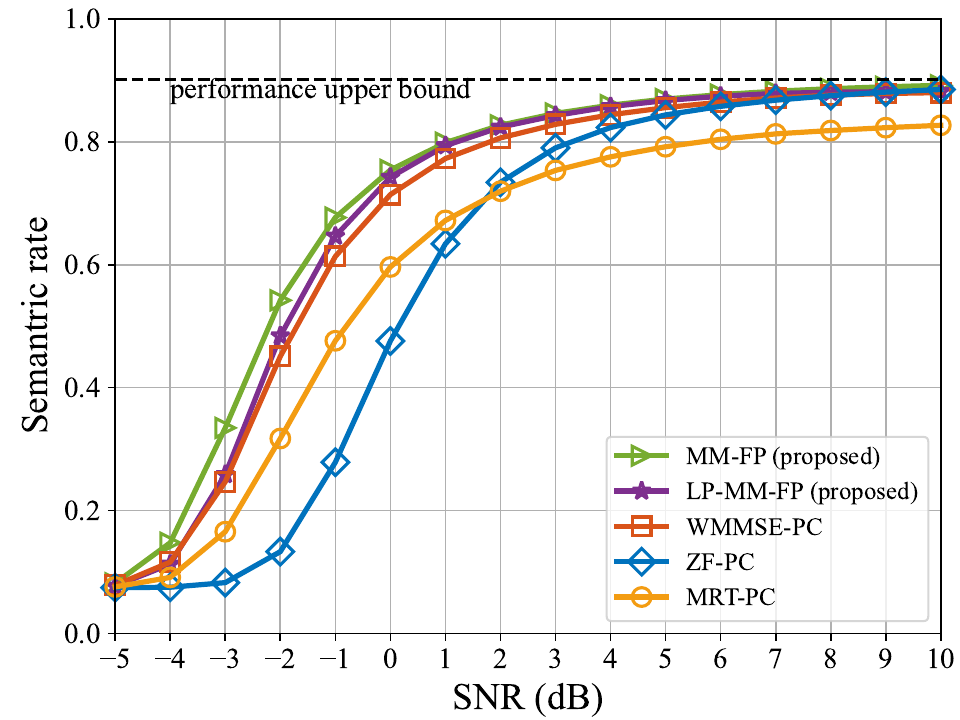}}
		\vspace{1mm}
			\caption{Performance comparison of different beamforming schemes}\label{Fig:beamform comparison}
		\vspace{-6mm}
	\end{figure}
	\subsection{Evaluation of Coexisting System} 
	\rv{ In this subsection, we evaluate the effectiveness of the semantic-bit coexisting system by comparing it with the BitCom system. The semantic-bit coexisting system utilizes JSCC with neural network for image transmission. As a benchmark, we consider the BitCom scheme, where BS employs the standard separate source and channel scheme to transmit images to sem-users. Specifically, BPG is used for source coding with a compression quality set to $34$. For channel coding, we adopt the Turbo codes following the LTE standard \cite{3gppturbo}, with a coding rate of $\frac{1}{3}$ and a block length of $2048$. For modulation, we utilize the $64$QAM scheme along with a soft demodulation process. The two different systems result in two sets of $\{a_K, d_K, c_K, e_K\}_{K=1}^L$, and we conduct Algorithm \ref{algo:sca fp algorithm} for beamforming under the two parameter sets.} 
	The results are presented in Fig. \ref{Fig:transmission system comparison}, where Fig. \ref{Fig:system comparison: low snr case} {shows} the performance {in} low SNR case (SNR = $0$dB), and Fig. \ref{Fig:system comparison: high snr case} {in} high SNR case (SNR = $5$dB).
\rv{In the low SNR case,  the BitCom system fails to work under any of {the examined} QoS requirements. 
}	This is because BitCom is sensitive to noise. 
	In the high SNR case, with a low QoS requirement, sem-users in the BitCom system enjoy low interference from bit-users, and thanks to the channel coding, the receiver is able to perfectly decode the BPG bit flow {and attains} good performance when $\beta_i \leq 0.5$. {With} the increase of QoS requirements, the strong interference causes the performance of BitCom to degrade {quickly}. 
    {In the meantime, 
    the semantic-bit coexisting system only experiences a slight performance degradation as the QoS requirement increases from 0 to 1.5}, 
    thus demonstrating its {effectiveness}. 

	\vspace{-3mm}
	Since data driven method is used to approximate the semantic rate, it is important to compare the real semantic rate with the approximated one. 
	We first implement the proposed beamforming scheme in Algorithm \ref{algo:sca fp algorithm} and then compare the image recovery quality (i.e., SSIM) and the objective value in $\mathbf{P1}$. The performance comparison is presented in Fig. \ref{fig: theoretical validation}, where three different $K$ settings are considered. 
    The final performance is averaged {over $2,000$ test samples}.
	As shown in Fig. \ref{fig: theoretical validation}, the approximation and simulation curves almost overlap, and the {approximating-based method }can well capture the performance growth trend as SNR increases. 
	This validates the effectiveness of the data driven method in accurately approximating the semantic rate. 
    \vspace{-2mm}
	\subsection{{Performance of Beamforming Design}} 
    \vspace{-1mm}
	In this subsection, we compare the performance of the proposed beamforming algorithm with three benchmark schemes. 
	Fig. \ref{Fig:beamform comparison} depicts the semantic performance using different beamforming schemes in the coexisting system. 
	We evaluate performance across different SNR and QoS settings. 
	Under different QoS settings, Fig. \ref{Fig:beamforming different qos}  shows that heuristic beamforming schemes, such as ZF-PC and MRT-PC, perform poorly {since they fail} to coordinate beamforming direction and power for the problem $\mathbf{P2}$. The optimization-induced method WMMSE-PC achieves better performance than ZF-PC and MRT-PC. However, WMMSE-PC {fails to consider the} semantic objective in Fig. \ref{Fig:performance on imagent} and (\ref{eq: the gt semantic rate function}), which has a different mapping relationship between SNR and performance. As a result, the performance of WMMSE-PC degrades significantly when QoS requirements increase, 
    {which implies that tailored design of beamforming for coexisting systems is required. }
    {The proposed beamforming algorithm outperforms the three benchmark schemes in all QoS settings, achieving the {best} performance {given all the examined} QoS requirements.
    } 
    Moreover, the proposed LP-MM-FP algorithm achieves near performance with MM-FP algorithm especially in low QoS regime, while with much lower computational complexity. 
    We also present some test examples in Fig. \ref{fig:example of beamform comparison different qos}, where $\beta_{i}$ is set to 0.8. The recovered image from the system that adopts the ZF-PC or MRT-PC beamforming schemes has an obvious blur, which is also reflected in SSIM performance. The system with the WMMSE-PC algorithm has relatively more noise points in the first and third image. The system with the proposed beamforming scheme recovers the first and second image clearly, with some blurs in the third image, yet still outperforms the other three benchmark schemes in terms of SSIM. 

	Fig. \ref{Fig:beamforming different snr} illustrates the performance comparison across different SNR settings. ZF-PC performs poorly in the low SNR regime, although it can approach the performance upper bound like WMMSE-PC and the proposed method {when ${\rm SNR\geq 6}$ dB}. MRT-PC performs relatively well in the low SNR regime, but {the performance quickly degrades compared with other schemes as SNR increases} since it does not consider user interference for beamforming design. Similarly, the proposed scheme outperforms these benchmark schemes in {all the examined} SNR settings, particularly in the low SNR regime, demonstrating its robustness. We present some test examples in Fig. \ref{fig:example of beamform comparison different snr}, and the recovery performance is consistent with the analytical results in Fig. \ref{Fig:beamforming different snr}. The system that adopts the proposed beamforming scheme achieves the best SSIM performance in all three recovered images.
	\vspace{-4mm}
	\subsection{\rv{Complexity Comparison of Beamforming Algorithms}}\label{subsec: complexity comparison}
		\begin{table}[h]
			\begin{tabular}{|l|l|l|l|l|l|}
			\hline
			\begin{tabular}[c]{@{}l@{}}\#of QoS and \\SNR ($\beta_i,{\rm SNR}$)\end{tabular} & \begin{tabular}[c]{@{}l@{}}MRT\\ -PC/ms\end{tabular} & \begin{tabular}[c]{@{}l@{}}ZF-PC\\/ms\end{tabular} & \begin{tabular}[c]{@{}l@{}}WMMSE\\ -PC/ms\end{tabular} & \begin{tabular}[c]{@{}l@{}}MM-FP\\ /ms\end{tabular} & \begin{tabular}[c]{@{}l@{}}LP-MM\\-FP/ms\end{tabular} \\ \hline
								($0.8,0$ dB)										   & 36.5                                              & 18.1                                             & 42.7                                                & 82.3                                                & 25.5                                                   \\ \hline
								($1.0,3$ dB)									   & 40.4                                              & 17.5                                             & 40.4                                                & 133.7                                               & 36.3                                                   \\ \hline
			\end{tabular}
			\vspace{3mm}
			\caption{\rv{The CPU Running Time of Beamforming Algorithms}}\label{table: complexity}
			\vspace{-4mm}
			\end{table}
			\rv{In this subsection, we evaluate the CPU execution time of various beamforming schemes on Intel I9-9900K CPU. As shown in Table \ref{table: complexity}, the proposed MM-FP algorithm has the highest computation time due to the iterative nature. By eliminating the need for iterative optimization, the proposed LP-MM-FP algorithm significantly reduces computation time. Specifically, LP-MM-FP has a similar CPU runtime to the MRT and ZF algorithms and is faster than the WMMSE algorithm. This indicates that the LP-MM-FP algorithm offers a complexity comparable to low-complexity methods like MRT and ZF while delivering competitive performance with the WMMSE algorithm, underscoring its practicality.} 
	\vspace{-3mm}
	\subsection{Evaluation of $K$ Configuring Strategies}\label{subsec: k evaluation}
	\begin{figure}[h!]
		\centering
		\includegraphics[width=0.8\linewidth]{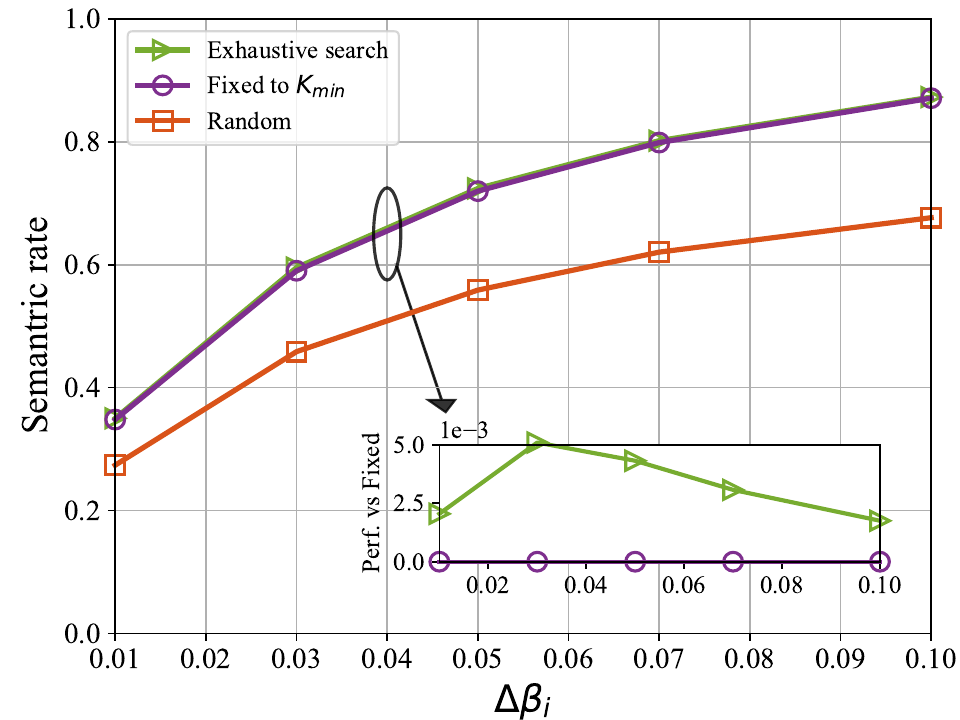}
		\vspace{1mm}
		\caption{{Performance comparison of different $K$-setting}}\label{fig: performance comparison of different k setting}
		\vspace{-4mm}
		\label{Fig:AD test acc}
	\end{figure}
	\begin{figure*}[htb]
		\begin{center}
			\resizebox{\textwidth}{!}
			{\begin{tabular}{cccccc}
	  {\Large Original} & {\Large ZF-PC} & {\Large MRT-PC} & {\Large WMMSE-PC} & {\Large LP-MM-FP}& {\Large MM-FP}\\
	  \includegraphics[width=0.18\textwidth]{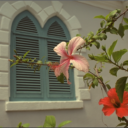}  &
	  \includegraphics[width=0.18\textwidth]{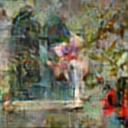}  &
	  \includegraphics[width=0.18\textwidth]{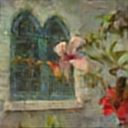}& 
	  \includegraphics[width=0.18\textwidth]{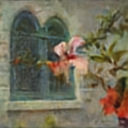}& 
	  \includegraphics[width=0.18\textwidth]{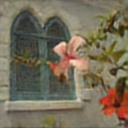}&
	  \includegraphics[width=0.18\textwidth]{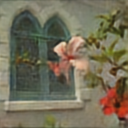}
	  \\
	  \Large SSIM & \Large 0.553 & \Large 0.699 & \Large 0.750 & \Large 0.781&\Large 0.803  \\
	  \includegraphics[width=0.18\textwidth]{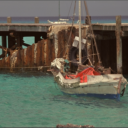}  &
	  \includegraphics[width=0.18\textwidth]{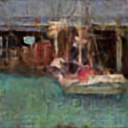}  &
	  \includegraphics[width=0.18\textwidth]{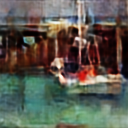}& 
	  \includegraphics[width=0.18\textwidth]{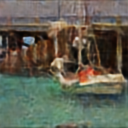}& 
	  \includegraphics[width=0.18\textwidth]{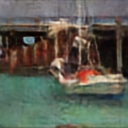}&
	  \includegraphics[width=0.18\textwidth]{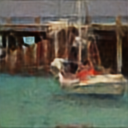}
	  \\
	  \Large SSIM & \Large 0.717 & \Large 0.642 & \Large 0.766 & \Large 0.812 & \Large 0.851\\  
	\end{tabular}}
	   \end{center}
		  \caption{\rv{Examples of reconstructed images under ${\rm SNR}=0$ dB, $\beta_{i}=0.8, \forall i =1,...,B$.}}
		  \label{fig:example of beamform comparison different qos}
		  \vspace{-3mm}
	  \end{figure*}
	  \begin{figure*}[htb]
		\begin{center}
			\resizebox{\textwidth}{!}
			{\begin{tabular}{cccccc}
	  {\Large Original} & {\Large ZF-PC} & {\Large MRT-PC} & {\Large WMMSE-PC} & {\Large LP-MM-FP} & {\Large MM-FP}\\
	  \includegraphics[width=0.18\textwidth]{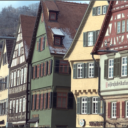}  &
	  \includegraphics[width=0.18\textwidth]{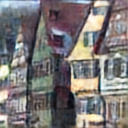}  &
	  \includegraphics[width=0.18\textwidth]{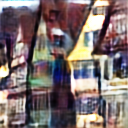}& 
	  \includegraphics[width=0.18\textwidth]{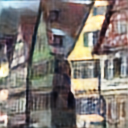}& 
	  \includegraphics[width=0.18\textwidth]{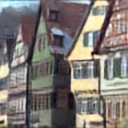}& 
	  \includegraphics[width=0.18\textwidth]{./Figures/beamforming_comparison_snr_3_qos_1.0/select/SCA_FP/1.png} 
	  \\
	  \Large SSIM & \Large 0.757 & \Large 0.691 & \Large 0.789 & \Large 0.831& \Large 0.848 \\
	  \includegraphics[width=0.18\textwidth]{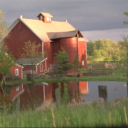}  &
	  \includegraphics[width=0.18\textwidth]{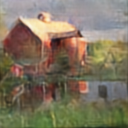}  &
	  \includegraphics[width=0.18\textwidth]{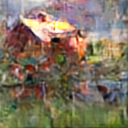}& 
	  \includegraphics[width=0.18\textwidth]{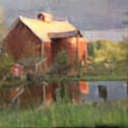}& 
	  \includegraphics[width=0.18\textwidth]{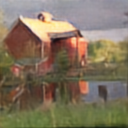} &
	  \includegraphics[width=0.18\textwidth]{./Figures/beamforming_comparison_snr_3_qos_1.0/select/SCA_FP/2.png} 
	  \\
	  \Large SSIM & \Large 0.781 & \Large 0.475 & \Large 0.827 & \Large 0.834 & \Large 0.844\\  
	\end{tabular}}
	   \end{center}
		  \caption{\rv{Examples of reconstructed images under SNR=3dB, $\beta_{b_i}=1.0, \forall i=1,...,B$.}}
		  \label{fig:example of beamform comparison different snr}
		  \vspace{-4mm}
	  \end{figure*}
	This subsection evaluates the effectiveness of {different methods for the configuration of $K$ in 
	a typical loaded scenario, i.e., $N_t=16,~B=14,~T=1$.
	} 
	We compare the exhaustive search against two benchmark schemes: Random, which randomly selects a value of $K$ from $1$ to $L$; the best fixed $K$ setting, which we found to be the minimum value $K=2$ based on numerical experiments.
	We present the performance comparison in Fig. \ref{fig: performance comparison of different k setting}, where we use $\Delta\beta_{i}$ to denote the gap between the currently selected QoS value and the maximum achievable QoS, and a smaller $\Delta\beta_{i}$ indicates a more stringent QoS requirement. We observe that the performance of all schemes improves as $\Delta\beta_{i}$ increases, with the Random algorithm performing noticeably worse than the other three schemes. 
    {The fixed $K$ algorithm has already achives satisfactory performance in the transmission task considered in this paper, which can be adopted in the scenarios with limited computation capability.  
	} 
    The considered method optimizes $K$ through the exhaustive method and can achieves the best performance.  
      \vspace{-4mm}
      \section{Conclusion}\label{sec: conclusion}
      \vspace{-1mm}
      In this paper, 
      {we considered a semantic-user and bit-user coexisting system. 
      A beamforming problem that maximizes the semantic rate under QoS constraints from bit-users and power constraint was formulated and solved in an low-complexity manner.} 
      Experiments show that the proposed method {significantly improves} the existing beamforming methods {dedicated} for BitCom. 
      {	Addressing issues beyond beamforming in the coexisting system remains an interesting future direction. }	  
	\bibliographystyle{ieeetr}
	\vspace{-3mm}
	\bibliography{BibDesk_File_v2}
	\vspace{-5mm}
	\end{document}